\documentclass[12pt,preprint]{aastex}
\usepackage{amsmath,natbib,psfig,emulateapj5}

\input{buote.defs}

\begin{document} 

\title{XMM-Newton and Chandra Observations of the Galaxy Group NGC
5044.\\ II. Metal Abundances and Supernova Fraction}
\author{David A. Buote\altaffilmark{1}, Aaron
D. Lewis\altaffilmark{1}, Fabrizio Brighenti\altaffilmark{2,3}, \&
William G. Mathews\altaffilmark{2}} 
\altaffiltext{1}{Department of Physics and Astronomy, University of
California at Irvine, 4129 Frederick Reines Hall, Irvine, CA 92697-4575}
\altaffiltext{2}{UCO/Lick Observatory, Board of Studies in Astronomy
and Astrophysics, University of California, Santa Cruz, CA 95064}
\altaffiltext{3}{Dipartimento di Astronomia, Universit\`a di Bologna,
via Ranzani 1, Bologna 40127, Italy}

\slugcomment{To Appear in The Astrophysical Journal}

\begin{abstract}
Using new \xmm\ and \chandra\ observations we present an analysis of
the metal abundances of the hot gas within a radius of 100~kpc of the
bright nearby galaxy group NGC 5044. Motivated by the inconsistent
abundance and temperature determinations obtained by different
observers for X-ray groups, we provide a detailed investigation of the
systematic errors on the derived abundances considering the effects of
the temperature distribution, calibration, plasma codes, bandwidth,
Galactic \nh, and background rate. The iron abundance (\fe) drops from
$\fe\approx 1\solar$ within $R\approx 50$~kpc to $\fe\approx 0.4$
solar near $R=100$~kpc.  This radial decline in \fe\ is highly
significant: $\fe=1.09\pm 0.04 \solar\, (\rm statistical)\, \pm
0.05\solar\, + 0.18\solar\, (\rm systematic)$ within $R=48$~kpc
($5\arcmin$) compared to $\fe=0.44\pm 0.02 \solar\, (\rm
statistical)\, \pm 0.10\solar\, + 0.13\solar\, (\rm systematic)$ over
$R=48-96$~kpc ($5\arcmin-10\arcmin$). There is no evidence that the
radial profile of \fe\ flattens at large radius. The data rule out
with high confidence a very sub-solar value for \fe\ within $R=48$~kpc
confirming that previous claims of very sub-solar central \fe\ values
in NGC 5044 were primarily the result of the Fe Bias: i.e., the
incorrect assumption of spatially isothermal and single-phase gas when
in fact temperature variations exist. Next to iron the data provide
the best constraints on the silicon and sulfur abundances. Within
$R=48$~kpc we obtain $\si/\fe = 0.83 \pm 0.02\, (\rm statistical)\,
\pm 0.02\, +0.07\, (\rm systematic)$ and $\su/\fe = 0.54 \pm 0.02\,
(\rm statistical)\, \pm 0.01\, +0.01\, (\rm systematic)$ in solar
units. These ratios (1) are consistent with their values at larger
radii, (2) strongly favor convective deflagration models over
delayed-detonation models of \snia\, and (3) imply that \snia\ have
contributed 70\%-80\% of the iron mass within a 100~kpc radius of NGC
5044. This \snia\ fraction is also similar to that inferred for the
Sun and therefore suggests a stellar initial mass function similar to
that of the Milky Way. We mention that at the very center ($R\approx
2$~kpc) the \xmm\ and \chandra\ CCDs and the \xmm\ RGS show that the
Fe and Si abundances drop to $\approx 50\%$ of their values at
immediately larger radius analogously to that seen in some galaxy
clusters observed with \chandra. We find the magnitude of this dip to
be sensitive to assumptions in the spectral model, but if real it is
difficult to reconcile with the expectation that metal enrichment from
the stars in the central galaxy should result in a centrally peaked
metal abundance profile in the hot gas.

\end{abstract}


\section{Introduction}
\label{intro}

There is presently a controversy associated with the iron abundances
of groups (and the most X-ray luminous elliptical galaxies) deduced
from X-ray observations. While there seems to be general agreement of
sub-solar iron abundances outside the central regions ($r\approx
50-100$~kpc) of groups \citep[e.g.,][]{fino99,buot00c}, different
investigators have often obtained (for the same groups) different
results for the central regions ($r\la 50$~kpc) where the metal
enrichment from a central galaxy should be most pronounced. Most
previous \rosat\ and \asca\ studies have found very sub-solar values
of \fe\ in the central regions of groups \citep[for reviews
see,][]{buot00a,mulc00}. Since these low values of
\fe\ are generally lower than the stellar iron abundances
\citep[e.g.,][]{trag00a}, they imply that Type~Ia supernovae (\snia)
cannot have contributed significantly to the enrichment of the hot
gas.  This implies that there is a lower binary star fraction and
\snia\ rate in the group galaxies so that most of the iron derives
from \snii\ with a ``top heavy'' stellar initial mass function (IMF)
\citep[e.g.,][]{renz93,renz97,arim97}. Consequently, various authors
have questioned the reliability of X-ray determinations of
\fe\ and have suggested that the low \fe\ values are caused by errors
associated with the Fe L lines in X-ray plasma codes
\citep[e.g.,][]{arim97,renz00}.

However, in a series of papers \citep{buot98c,buot99a,buot00a,buot00c}
we found that indeed the iron abundances in the central regions of
groups were measured incorrectly, but not because of errors in the
plasma codes. Instead, we attributed the very sub-solar \fe\ values to
an ``Fe Bias'' arising from forcing a single-temperature model to fit
a spectrum consisting of multiple temperature components with
temperatures near 1 keV \citep[see
especially,][]{buot00a,buot00c}. The multiple temperature components
may arise either from a radially varying single-phase gas or represent
real multiphase structure in the hot gas. We found near-solar values
for \fe\ within the central 50-100~kpc of groups, which is larger than
the typical stellar value for \fe\ within $R_e/2$ in elliptical
galaxies \citep{trag00a}, implying that a significant number of \snia\
have enriched the hot gas, in better agreement with a Galactic IMF.

Even stronger constraints on the \snia\ fraction and the IMF are
placed by the ratios of the abundances of $\alpha$ elements to iron
\citep[e.g.,][]{gibs97,renz97,brig99a}. Previous X-ray observations did not
place strong constraints on the $\alpha$ abundances in groups and were
usually consistent with solar values \citep[e.g.,][and references
therein]{mulc00}.

Recently, using a new \xmm\ observation of the luminous X-ray group
NGC 1399 we find $\fe/\solar\approx 1.5-2$ within $r\approx 20$~kpc and
$\si/\fe\approx 0.8$ solar over the entire 50~kpc radius
studied. These results imply a \snia\ fraction of $\approx 80\%$ which
is similar to that inferred for the Sun and therefore suggests a
stellar initial mass function similar to the Milky Way as advocated by
Renzini and others \citep[e.g.,][]{renz93,renz97,wyse97}.

NGC 5044 is brighter and more luminous in X-rays than NGC 1399 but is
slightly lower in temperature. In \citet[][hereafter Paper 1]{buot03a}
we showed using \xmm\ and \chandra\ data that within $r\approx 30$~kpc
the hot gas is not isothermal, nor is it consistent with a radially
varying single-phase medium. Instead a limited multiphase medium is
implied where the temperature varies from $\sim \thot$ to $\sim
\thot/2$ ($\thot\approx 1.3$~keV), but no lower.

In this paper we measure the metal abundances of the hot gas in NGC
5044 using the \xmm\ and \chandra\ data and provide a detailed
investigation of the systematic errors in the derived abundances
considering the effects of the temperature distribution, calibration,
plasma codes, bandwidth, Galactic \nh, and background rate. The
implications of these measurements for the supernova fraction and IMF
are then discussed.

The paper is organized as follows. In \S \ref{fe} we present the iron
abundance as a function of radius for different spectral models. We
present the profiles of silicon in \S \ref{si} and other abundances in
\S \ref{other}. A comprehensive discussion of systematic errors in the
abundance measurements is given in \S \ref{sys}.  We give in \S
\ref{bias} our most precise constraints for the emission-weighted
average abundances in regions of 0-50~kpc and 50-100~kpc with a full
and explicit accounting of the relevant systematic errors. Finally, in
\S \ref{conc} we present our conclusions. We assume a distance to NGC
5044 of 33~Mpc using the results of \citet{tonr01} for
$H_0=70$~\kmsmpc\ (note: $1\arcsec=0.160$~kpc).

\section{Iron Abundance}
\label{fe}

\begin{figure*}[t]
\parbox{0.49\textwidth}{
\centerline{\psfig{figure=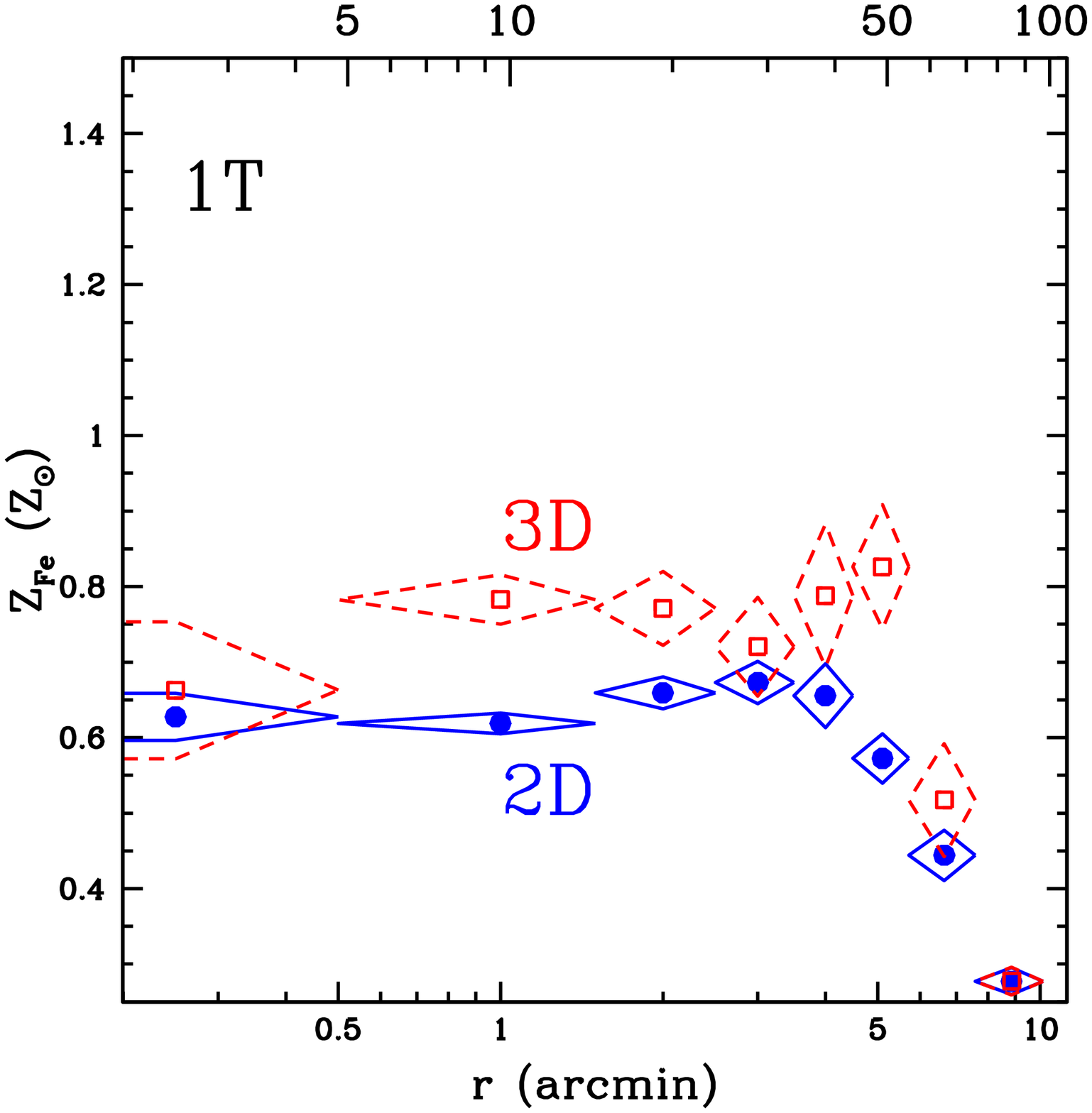,height=0.3\textheight}}}
\parbox{0.49\textwidth}{
\centerline{\psfig{figure=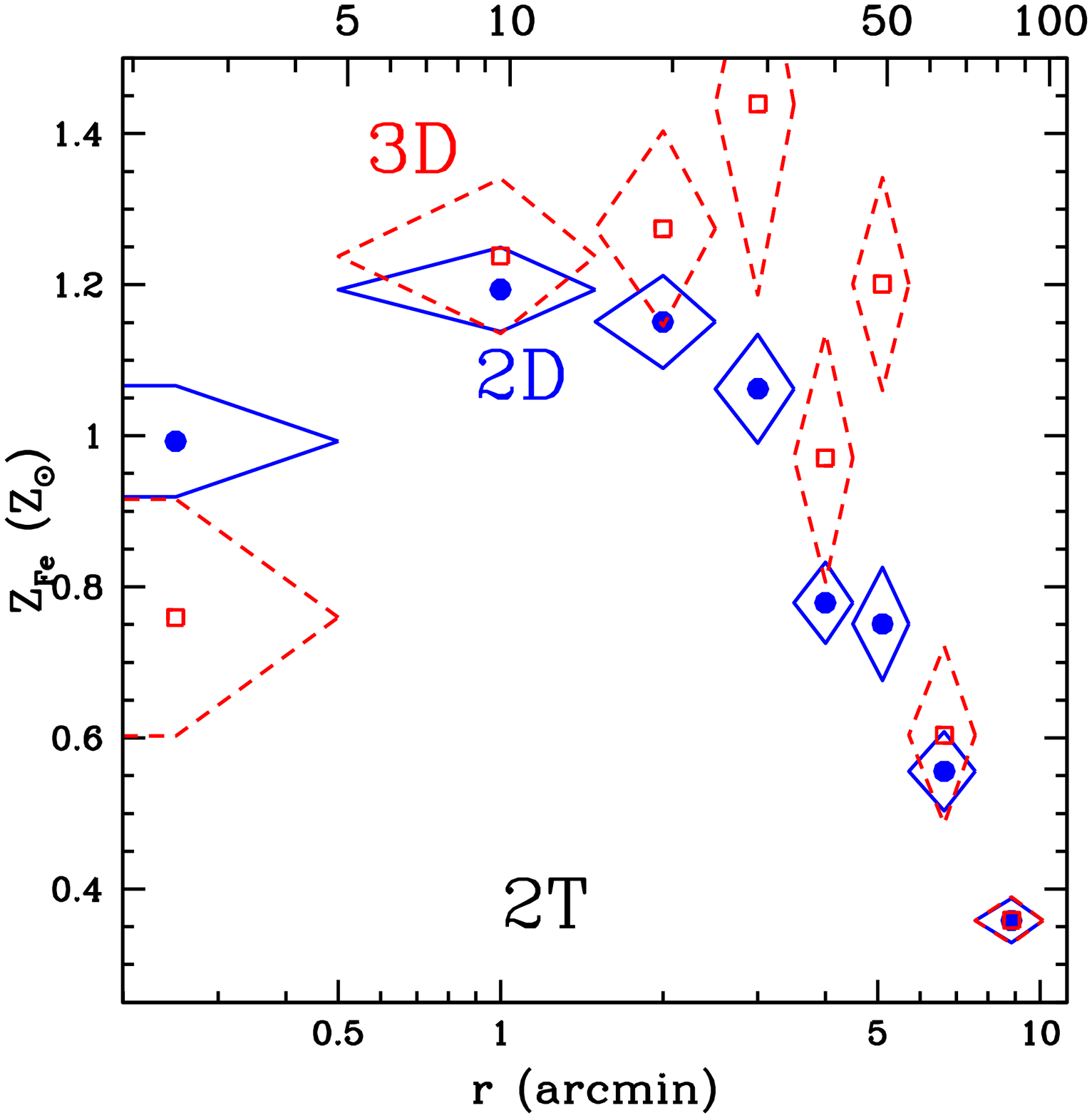,height=0.3\textheight}}}
\caption{\label{fig.fe} Radial profiles
(units -- bottom: arcminutes, top: kpc) of the Fe abundance and
associated $1\sigma$ errors for ({\sl Left panel}) 1T and ({\sl Right
panel}) 2T models fitted simultaneously to the \xmm\ and \chandra\
data. Note that the \chandra\ data apply only to the inner three
radial bins.  In each case ``3D'' refers to results obtained from a
spectral deprojection analysis of the \xmm\ and \chandra\ data.}
\end{figure*} 

\begin{figure*}[t]
\centerline{\psfig{figure=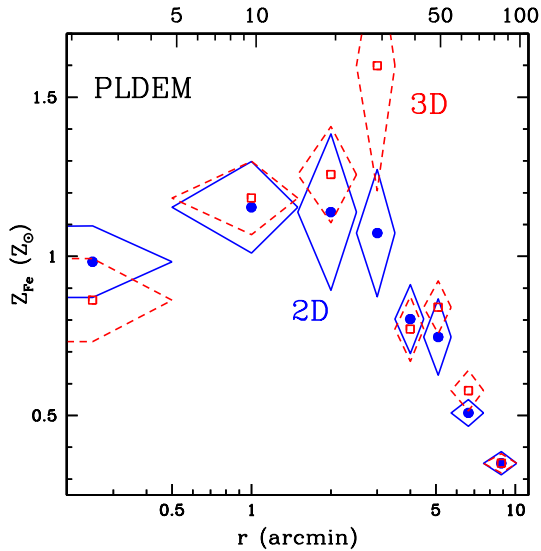,height=0.3\textheight}}
\caption{\label{fig.pldem} Same as Figure \ref{fig.fe} except for the
model with a power-law DEM (PLDEM).}
\end{figure*} 

\begin{table*}[t] \footnotesize
\caption{Temperatures for 1T and 2T Models
\label{tab.temps}} 
\begin{center} \vskip -0.4cm
\begin{tabular}{ccc|c|c|cc|cc} \tableline\tableline\\[-7pt]
&  &  & 1T (2D) & 1T (3D) & \multicolumn{2}{c}{2T (2D)} & \multicolumn{2}{c}{2T (3D)}\\
& $R_{\rm in}$ & $R_{\rm out}$ & $T$ & $T$ & \tcool & \thot & \tcool & \thot\\
Bin & (arcmin) & (arcmin) & (keV) & (keV) & (keV) & (keV) & (keV) & (keV)\\ 
\tableline \\[-7pt]
1  & 0.  & 0.5  & $0.754\pm 0.005$  & $0.698\pm 0.013$  & $0.74\pm 0.01$   & $1.66\pm 0.09$ & $0.69\pm 0.01$   & $3.7(>1.4)$   \\
2  & 0.5 & 1.5  & $0.807\pm 0.002$  & $0.788\pm 0.003$  & $0.787\pm 0.003$ & $1.47\pm 0.04$ & $0.772\pm 0.004$ & $1.62\pm 0.07$\\
3  & 1.5 & 2.5  & $0.985\pm 0.003$  & $0.904\pm 0.006$  & $0.83\pm 0.01$   & $1.35\pm 0.02$ & $0.82\pm 0.01$   & $1.36\pm 0.06$\\
4  & 2.5 & 3.5  & $1.172\pm 0.005$  & $1.111\pm 0.018$  & $0.87\pm 0.01$   & $1.36\pm 0.02$ & $0.86\pm 0.01$   & $1.37\pm 0.06$\\
5  & 3.5 & 4.5  & $1.236\pm 0.010$  & $1.251\pm 0.019$  & $0.92\pm 0.01$   & $1.31\pm 0.02$ & $1.05\pm 0.08$   & $1.47\pm 0.13$\\
6  & 4.5 & 5.7  & $1.229\pm 0.007$  & $1.244\pm 0.017$  & $0.85\pm 0.10$   & $1.35\pm 0.03$ & $0.86\pm 0.14$   & $1.37\pm 0.04$\\
7  & 5.7 & 7.6  & $1.205\pm 0.011$  & $1.268\pm 0.025$  & $0.79\pm 0.14$   & $1.29\pm 0.05$ & $0.50\pm 0.23$   & $1.28\pm 0.05$\\
8  & 7.6 & 10.1 & $1.147\pm 0.013$  & $1.147\pm 0.015$  & $0.90\pm 0.07$   & $1.30\pm 0.04$ & $0.90\pm 0.10$   & $1.30\pm 0.05$\\
\tableline \\[-1.0cm]
\end{tabular}
\tablecomments{Temperature results obtained in Paper 1  for 1T
and 2T models. These models have been fitted simultaneously to the
\xmm\ and \chandra\ data in bins 1-3 and only to the \xmm\ data for bins
4-8.  ``Bin'' refers either to a circular annulus (2D) or spherical
shell (3D). For the case where a lower limit is given it represents
the lowest value obtained from 100 error simulations; i.e., it is
essentially a 99\% confidence lower limit.}
\end{center}
\end{table*}

\subsection{Preliminaries}

To obtain the three-dimensional properties of the X-ray emitting gas
we perform a spectral deprojection analysis assuming spherical
symmetry using the (non-parametric) ``onion-peeling'' technique as
discussed in Paper 1. We refer to deprojected models as ``3D'' while
traditional model fits to the data on the sky are referred to as
``2D'' (i.e., without deprojection). However, with respect to 2D
models, deprojection always inflates the errors between points which
is related to the error associated with the derivative of the
emissivity in an Abel inversion.  As discussed in Paper 1, we perform
a regularization procedure on the oxygen and neon abundances to limit
their radial fluctuations in 3D models. No regularization is applied
to any 2D model. Because the abundances obtained from 2D models have
smaller statistical errors and do not require any regularization we
shall generally present results for both 2D and 3D models in this
paper. Statistical errors on the parameters are computed using the
Monte Carlo procedure discussed in \S 4.1 of Paper 1.

We measure the Fe abundance as a function of radius using a suite of
different models for the temperature distribution as described in
Paper 1. Single-temperature (1T) and two-temperature (2T) models are
used as our baseline models for comparison to previous studies.  We
also examined a set of models that emit over a continuous range of
temperatures; i.e., models having a continuous differential emission
measure (DEM): cooling flow, gaussian DEM, and power-law DEM (PLDEM).
In every model the following abundances are free parameters: Fe, O,
Ne, Mg, Si, and S -- the abundances for all other elements are tied to
Fe in their solar ratios. Unless stated otherwise, for every
multitemperature model (e.g., 2T) the abundances of one temperature
component are tied to those in the other temperature component(s).
Even though the 2T and PLDEM models provide superior fits within the
central $\sim 30$~kpc of NGC 5044, in this paper we present results
for all models for the purpose of showing the dependence of the
inferred Fe abundance on the temperature structure of the hot gas.

The reference solar Fe abundance has been a source of much confusion
in the literature. There is now good agreement between values of the
Fe abundance obtained from measurements in the solar photosphere and
from solar-system meteorites
\citep[e.g.,][]{mcwi97,grsa}. Therefore, we take the solar abundances
in \xspec\ (v11.2.0af) to be those given by the
\citet{grsa} table which use the correct Fe value, $\rm Fe/H=3.2\times
10^{-5}$ by number.  Unfortunately, most previous and many current
X-ray studies of Fe abundances use the incorrect ``old photospheric''
value of $\rm Fe/H=4.7\times 10^{-5}$ present in the \citet{angr}
table in \xspec\ which is still the default option in
\xspec. Consequently, investigators who use the old photospheric value
for Fe/H obtain values for the Fe abundance that are approximately a
factor 1.4 too small. In comparing with our results, we shall
transform all abundances from previous studies to those of
\citet{grsa} unless stated otherwise.

\subsection{Results}

The Fe abundance (\fe) as a function of radius obtained from 1T models
fitted simultaneously to the \xmm\ and \chandra\ data is displayed in
Figure \ref{fig.fe} (Left panel); note that the \chandra\ data only
apply to the inner three radial bins. In Table \ref{tab.temps}, for
convenience, we reproduce the temperature results for the 1T and 2T
models from Paper 1. For $r\la 50$~kpc the iron abundance is
approximately constant such that $\fe\approx 0.65\solar$ for the 1T
(2D) model and $\fe\approx 0.75\solar$ for the 1T (3D) model. At
larger radii \fe\ decreases with increasing radius with the lowest
value occurring in the bounding annulus/shell: $\fe=0.28\pm
0.02\solar$ for $r=73-97$~kpc. The $\sim 15\%$ larger values of \fe\
obtained for the deprojected models near 40-50~kpc are primarily the
result of the projection effect which tends to smear out a radially
varying function, while for smaller radii they are primarily the
result of the Fe Bias. In the latter case the deprojection removes the
projected temperature components from exterior shells thus allowing
the 1T model to be a better -- though still not good -- representation
of the spectrum within a given shell at smaller radii.

Previous \asca\ studies of NGC 5044 that fitted 1T models to the
$\approx 0.5-10$~keV data in the central regions obtained much smaller
abundances. For example, \citet{fuka96} obtained $\fe=0.46\pm
0.03\solar$ for $R<20$~kpc, and \citet{fino99} obtained $\fe=0.65\pm
0.07\solar$ for $r=30-40$~kpc using 1T models of the \asca\ data of
NGC 5044. As we discussed in \citet{buot99a}, 1T models are poor fits
to the \asca\ data extracted from large apertures because they contain
a distribution of temperatures implied by the radial temperature
gradients observed with \rosat\ \citep{davi94,buot99a,buot00c}. These
temperature gradients suggested by \rosat\ are confirmed and mapped in
detail with \chandra\ and \xmm\ in Paper 1.

The Fe abundances for the 2T models fitted simultaneously to the \xmm\
and \chandra\ data are also displayed in Figure \ref{fig.fe} (Right
panel). For all radial bins interior to a radius of $\approx 40$~kpc,
excluding the central bin, the values of \fe\ obtained from the 2T
models exceed by 40\%-100\% those obtained from the 1T models. This
{\it systematic} increase arises from the Fe Bias, and the differences
between the 1T and 2T models are highly significant; e.g., in shell
\#2 we obtain $\fe=0.78\pm 0.03\solar$ for 1T (3D) $\fe=1.24\pm
0.10\solar$ for 2T (3D), and $\fe=1.18\pm 0.12\solar$ for PLDEM (3D)
models. Recall from Paper 1 that the 2T and PLDEM models provide the
best fits to the data, and the fit residuals observed for the 1T model
near 1~keV are fully consistent with those produced by the Fe
Bias.

In Figure \ref{fig.pldem} we show the Fe abundances obtained for the
PLDEM model.  It is notable that the PLDEM model, which fits the data
almost as well as the 2T model, gives \fe\ values consistent with the
2T model everywhere within the $1\sigma-2\sigma$ errors. The overall
similarity of \fe\ values for different multitemperature models
(including cooling flow and Gaussian DEM -- not shown) indicates that
the value of \fe\ can be fairly reliably inferred with any of the
multitemperature models; i.e., it is principally the 1T model that
suffers from the Fe Bias and gives large underestimates of \fe.

Focusing on the multitemperature models since they provide the best
spectral fits in the central regions, we conclude that the iron
abundance displays a strong gradient in NGC 5044. At the largest
radius examined ($R=97$~kpc) we have $\fe\approx 0.35\solar$. The
(deprojected) iron abundance increases to values between 1-1.5 solar
within $r\approx 50$~kpc for all multitemperature models and then dips
within the central bin to a value of $\approx 0.8\solar$.

For every model we have discussed so far (except the 1T (2D) model)
the iron abundance dips in the central bin. In \S \ref{conc} we
discuss a multi-temperature model that does not dip in the center and
the implications of such a dip for enrichment models of the hot gas.

\section{Silicon Abundance}
\label{si}

\begin{figure*}[t]
\parbox{0.33\textwidth}{
\centerline{\psfig{figure=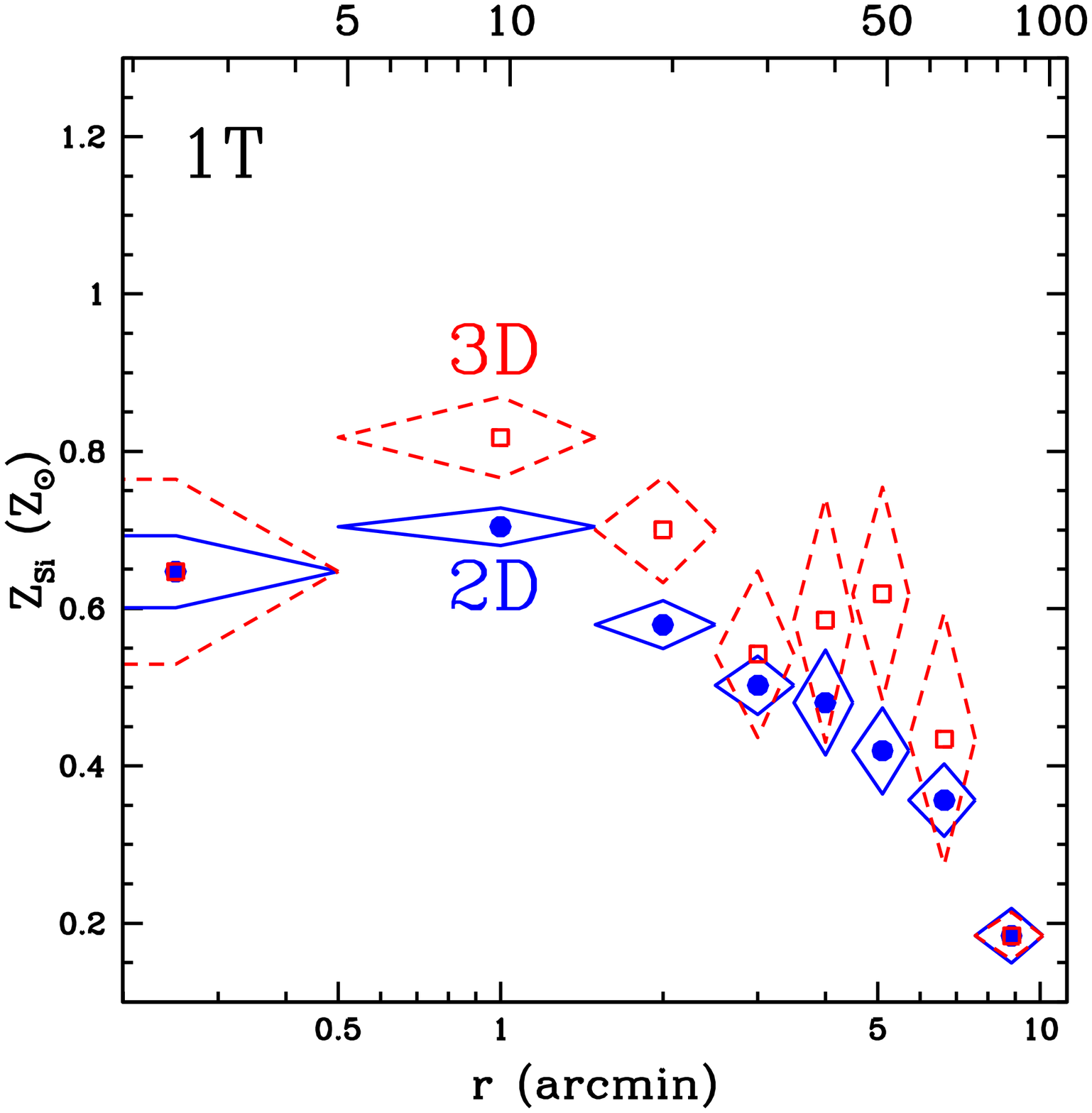,height=0.26\textheight}}
}
\parbox{0.33\textwidth}{
\centerline{\psfig{figure=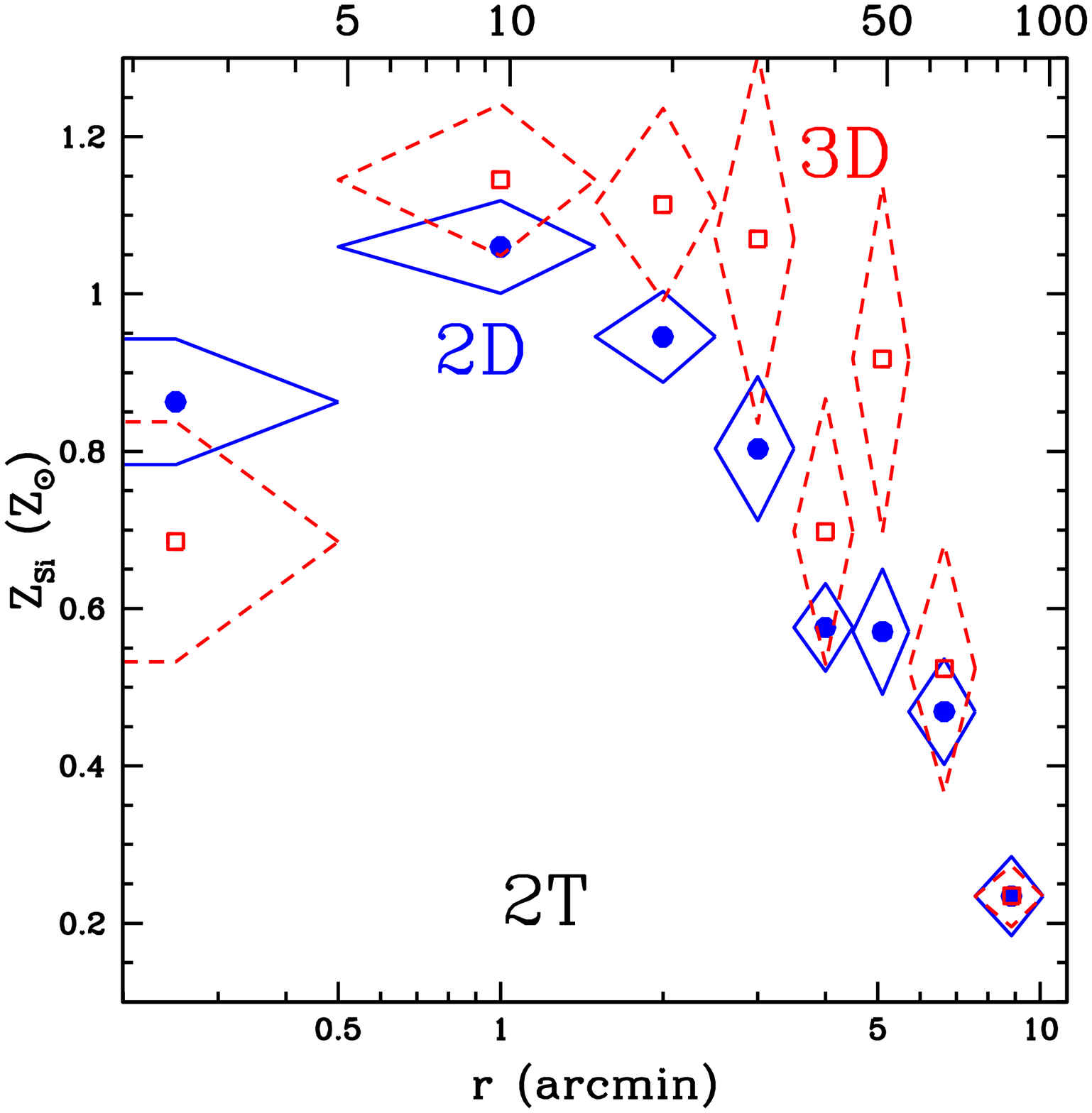,height=0.26\textheight}}
}
\parbox{0.33\textwidth}{
\centerline{\psfig{figure=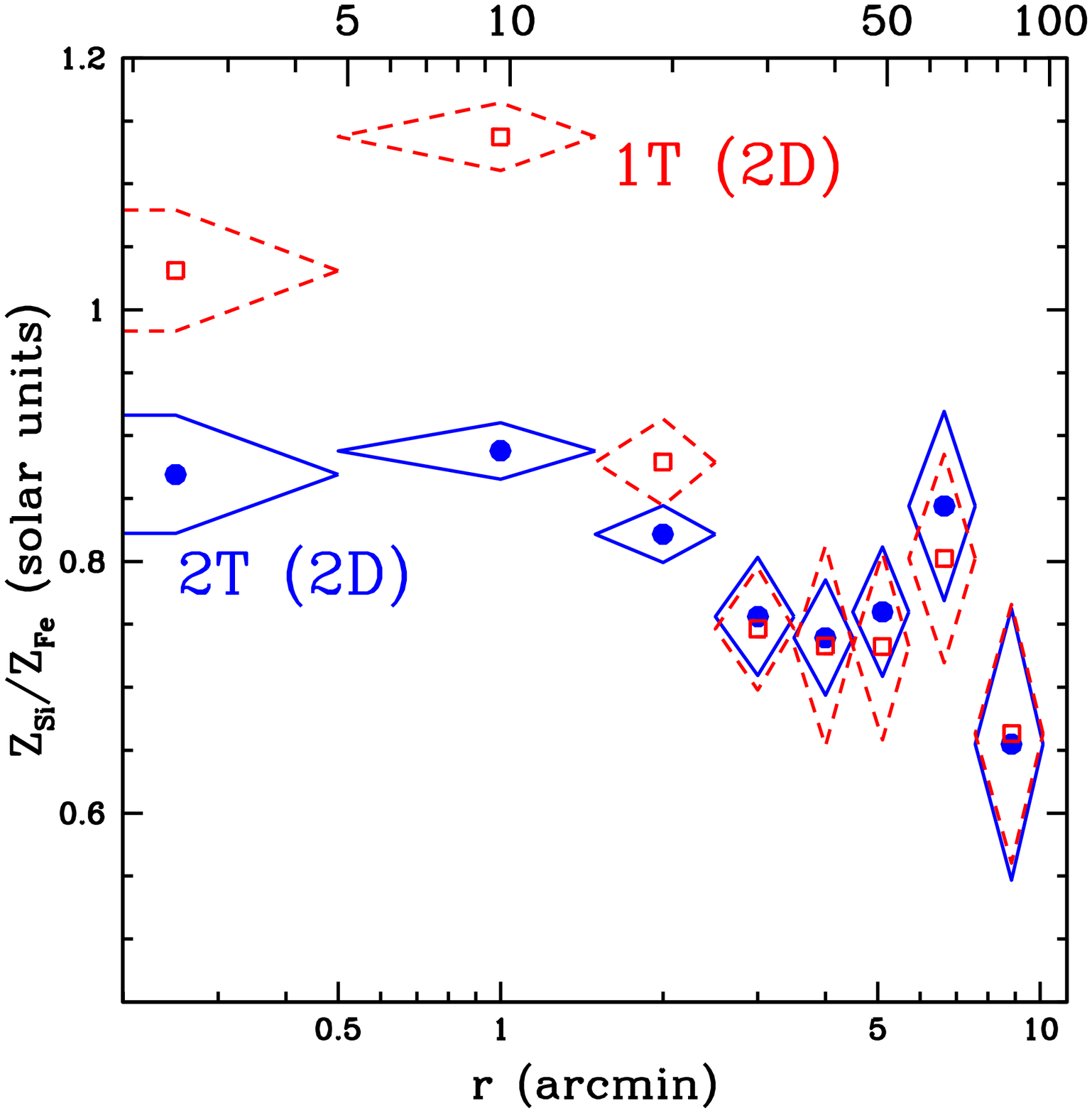,height=0.26\textheight}}
}
\caption{\label{fig.si} Radial profiles
(units -- bottom: arcminutes, top: kpc) of the Si abundance and
associated $1\sigma$ errors for ({\sl Left panel}) 1T and ({\sl Middle
panel}) 2T models. In each case ``3D'' refers to results obtained from
a spectral deprojection analysis. ({\sl Right panel}) \si/\fe\ ratios
for the 1T (2D) and 2T (2D) models.}
\end{figure*} 

Next to the broad feature of Fe L lines near 1 keV, the most notable
emission lines in the EPIC spectra of NGC 5044 (see Paper 1) are the
K$\alpha$ lines of silicon; i.e., \ion{Si}{13} \hea\ (1.85~keV),
\ion{Si}{14} \lya\ (2.0~keV). It follows that the silicon abundance
(\si) is also best constrained next to \fe. Although it might be
expected that \si\ does not suffer the same model-dependences as \fe\
because the silicon lines are fairly isolated and well-separated from
the Fe L lines, the results of the spectral fits show otherwise.

In Figure \ref{fig.si} we display \si\ as a function of radius for 1T
and 2T models. Analogously to \fe\ we find that \si\ obtained from
deprojected fits (i.e., 3D) of both 1T and 2T models are
systematically larger than those obtained from 2D fits by
10\%-20\%. For 1T models, $\si\approx 0.7\solar$ for $r\la 30$~kpc and
falls to $\si\approx 0.2\solar$ in the outermost radial bin. Similar
values are obtained at large radius for 2T models. But for $r\la
30$~kpc, 2T (3D) model yields $\si\approx 1.1\solar$ for $r\la 30$~kpc
which dips to $\si\approx 0.7\solar$ in the central bin in good
agreement with the behavior observed for \fe.

The other multitemperature models (cooling flow, Gaussian DEM, power
law DEM) give values for \si\ relative to the 1T and 2T models that
are entirely analogous to that described for \fe\ in the previous
section; i.e., they mostly give values consistent with the 2T model
within the 1-2 $\sigma$ errors. Given that the 2T and PLDEM models
provide the best fits within $r\la 30$~kpc (Paper 1), and they yield
fully consistent values for \fe\ and \si, the near-solar values
obtained for each element should be considered the favored values.

Although the silicon abundance obtained from 1T models within the
central regions is biased low because of multiple temperature
components having values near 1~keV within each radial bin
\citep[i.e., ``Silicon Bias'',][]{buot00a}, the ratio of silicon-to-fe
(\si/\fe) is affected much less. In Figure \ref{fig.si} (Right panel)
we plot \si/\fe\ as function of radius for 1T (2D) and 2T (2D) models;
the 3D models are everywhere consistent with their 2D counterparts
within the $1\sigma-2\sigma$ errors. The Gaussian-weighted mean of all
radial bins is $\si/\fe = 0.96\pm 0.02$ (in solar units) for 1T (2D)
and $\si/\fe = 0.83\pm 0.01$ (in solar units) for 2T (2D); the
gas-mass-weighted values are $\si/\fe = 0.74\pm 0.05$ and $\si/\fe =
0.74\pm 0.05$ respectively for the 1T and 2T models. As expected, the
largest differences between 1T and 2T models are observed within the
central four bins (i.e., $r\la 30$~kpc) where $\si/\fe = 1.00\pm 0.02$
and $\si/\fe = 0.85\pm 0.01$ respectively for the 1T and 2T models for
Gaussian means and $\si/\fe = 0.86\pm 0.03$ and $\si/\fe = 0.80\pm
0.02$ respectively for corresponding mass-weighted means; for
comparison, in the outer bins (5-8) we have $\si/\fe = 0.74\pm 0.04$
(1T) and $\si/\fe = 0.76\pm 0.03$ (2T) for Gaussian means and $\si/\fe
= 0.72\pm 0.06$ (1T) and $\si/\fe = 0.73\pm 0.06$ (2T) for
mass-weighted means.

We note that the marginal evidence that $\si/\fe$ decreases with
radius in NGC 5044 (i.e., for the 2T model the Gaussian mean value of
$\si/\fe$ for bins 5-8 is lower than that obtained from bins 1-4 by
$1.9\sigma$) should be considered quite tentative because this
relatively small difference could be an artifact of the simple
spectral models used. (However, all models we have investigated give
similar results.) If real, the slightly larger $\alpha$/Fe abundance
ratios near the center might be the result of stellar mass loss which
would be expected to have a lower \snia\ fraction.

\section{Other Abundances}
\label{other}

\begin{figure*}[t]
\parbox{0.49\textwidth}{
\centerline{\psfig{figure=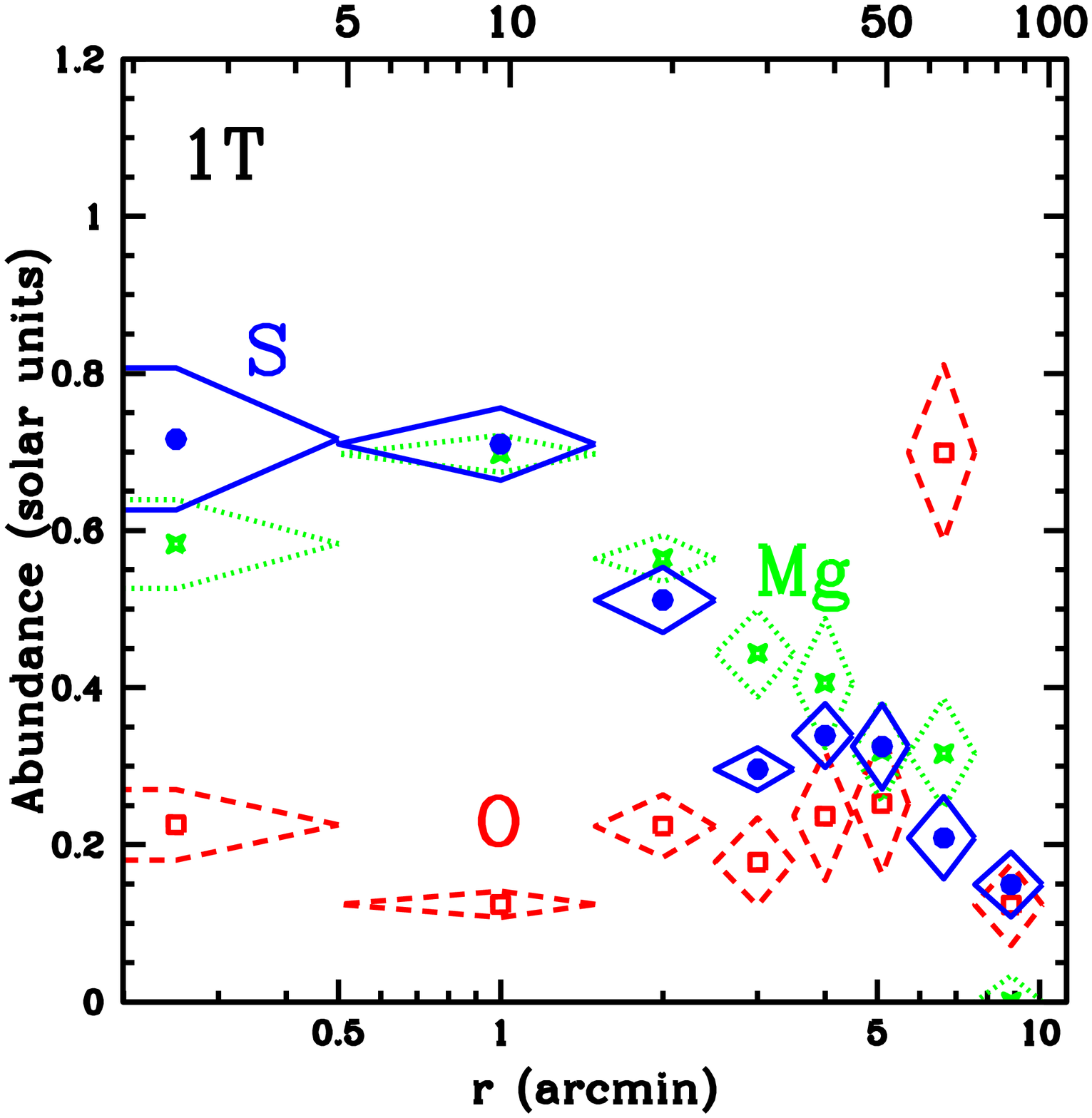,height=0.3\textheight}}
}
\parbox{0.49\textwidth}{
\centerline{\psfig{figure=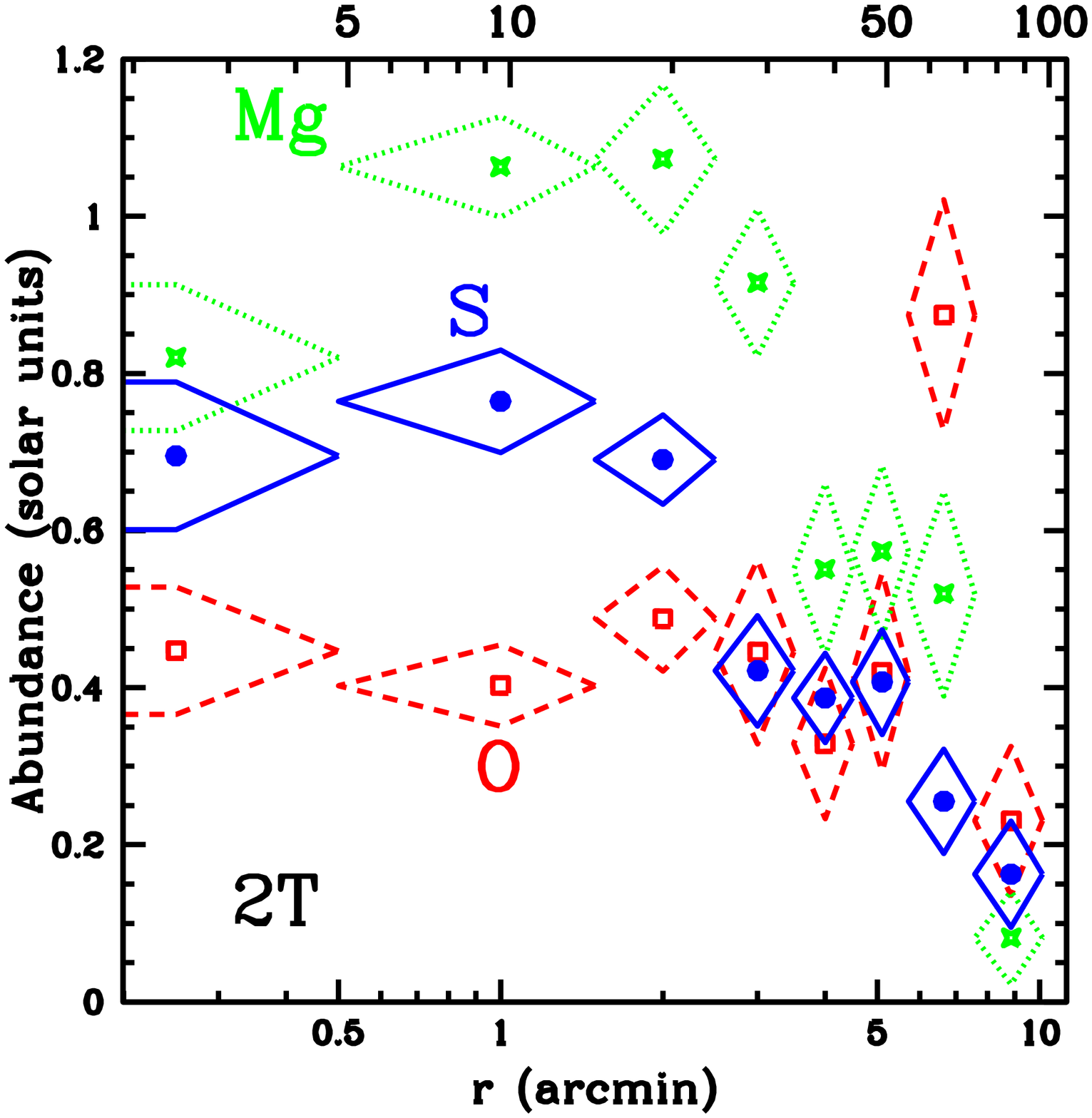,height=0.3\textheight}}
}

\caption{\label{fig.omg} Radial profiles
(units -- bottom: arcminutes, top: kpc) of the O, Mg, and Si abundances
and their associated $1\sigma$ errors for the ({\sl Left panel}) 1T (2D)
and ({\sl Right panel}) 2T (2D) models.}
\end{figure*} 

In Figure \ref{fig.omg} we show the radial abundance profiles for O,
Mg, and S for the 1T (2D) and 2T (2D) models; the deprojected values
(not shown) are considerably noisier and are in most cases consistent
with the 2D values within the $1\sigma$ errors. Although there is
considerable scatter between radii, like Si the values of the O and Mg
abundances obtained with the 2T model are systematically larger than
those obtained with the 1T model within $r\sim 50$~kpc. The values of
S are generally consistent between the 1T and 2T models. The relative
insensitivity of S to the temperature model is likely attributed to
the fact that the key S emission lines near 2.4~keV are farther away
from the $\sim 1$~keV Fe L lines than either O, Mg, or Si. 

Although the value of the Mg abundances differs significantly between
1T and 2T models, the Mg/Fe ratio -- like the Si/Fe ratio -- is quite
similar for both models; e.g., in bin \#2, $\mg/\fe=0.94\pm 0.05$ in
solar units, for 1T (3D) while, $\mg/\fe=0.88\pm 0.05$ in solar units,
for 2T (3D). But the O/Fe and S/Fe ratios do show significant
differences between the models, although the O/Fe ratio is very
sub-solar in both cases; e.g., in bin \#2, $\ox/\fe=0.19\pm 0.04$ and
$\su/\fe=0.96\pm 0.08$ in solar units, for 1T (3D) while,
$\ox/\fe=0.30\pm 0.05$ and $\su/\fe=0.66\pm 0.05$ in solar units, for
2T (3D). We note that the oxygen abundance in bin \#7 is significantly
larger ($\approx 4\sigma$) than its values in adjacent bins. We do not
believe the fluctuation in the oxygen abundance in bin \#7 is
physical, but we have not yet isolated the cause.

For the 2T (3D) model we obtain Gaussian weighted mean values of all
radial bins of $\su/\fe = 0.62\pm 0.04$, $\ox/\fe=0.39\pm 0.03$, and
$\mg/\fe=0.88\pm 0.04$ in solar units; the corresponding
gas-mass-weighted means are $\su/\fe = 0.46\pm 0.10$, $\ox/\fe=0.56\pm
0.09$, and $\mg/\fe=0.66\pm 0.17$ in solar units. Ignoring the
anomalous bin \#7 we obtain $\ox/\fe=0.37\pm 0.03$ (Gaussian averaged)
and $\ox/\fe=0.37\pm 0.04$ (mass averaged) including only bins
1-6. 

Finally, we mention that the Ne abundance (not shown) is the most
poorly constrained abundance that we have investigated. Undoubtedly
the poor constraints are at least partially attributed to the location
of the strongest Ne lines near 1~keV which are therefore completely
blended with the Fe L lines in the CCD spectra. Generally, we find
very sub-solar values for the Ne abundance except within $r\sim
10$~kpc where near solar values are suggested. More precise
constraints are provided for the emission-weighted average Ne
abundance in \S \ref{bias}. 

\section{Non-Azimuthally Symmetric Analysis}
\label{2d}

Following our discussion in \S 5 of Paper 1 we searched for azimuthal
variations in the iron abundance by fitting 1T and 2T models to
spectra extracted from circular apertures arranged in a square array
on the EPIC MOS images. Consistent with the results for the
temperatures discussed in Paper 1, we find that overall the abundances
obtained from this analysis are consistent with the spherically
symmetric analysis. In particular, we find no evidence for azimuthal
abundance variations associated with the small azimuthal temperature
variations between radii of $2\arcmin-3\arcmin$ mentioned in Paper 1.
We also observe the same increase in the iron abundance for apertures
within $r\sim 30$~kpc for 2T over 1T models that is obtained in the
spherically symmetric (i.e., azimuthally averaged) analysis.

\section{Systematic Errors}
\label{sys}

This section contains a detailed investigation of systematic errors on
the abundance measurements. Those readers who are not interested in
these technical details can safely skip ahead to \S \ref{bias}.

\subsection{Calibration}
\label{calib}

\begin{table*}[t] \footnotesize
\caption{Comparison of Fe Abundances from Chandra and XMM
\label{tab.chan}} 
\begin{center} \vskip -0.4cm
\begin{tabular}{cccc|cccccc} \tableline\tableline\\[-7pt]
& \multicolumn{3}{c}{1T} & \multicolumn{6}{c}{2T}\\ \tableline \\[-7pt]
& XMM & Chandra & \% & \multicolumn{2}{c}{XMM} & \multicolumn{2}{c}{Chandra} & \% & \% \\
Annulus & (free) & (free) & (free) & (free) & (fix) & (free) & (fix) &
 (free) & (fix)\\
\tableline \\[-7pt]
1 & $0.64\pm 0.05$ & $0.61\pm 0.05$ & $5\pm 1$  & $1.17\pm 0.13$ & $1.04\pm 0.13$  & $0.83\pm 0.09$ & $0.90\pm 0.09$ & $29\pm 14$ & $13\pm 16$ \\
2 & $0.58\pm 0.02$ & $0.68\pm 0.03$ & $-17\pm 6$ & $1.27\pm 0.07$ & $1.20\pm 0.07$  & $1.11\pm 0.07$ & $1.20\pm 0.07$ & $13\pm 8$ & $0\pm 8$ \\
3 & $0.65\pm 0.03$ & $0.67\pm 0.03$ & $-3\pm 6$  & $1.30\pm 0.07$ & $1.24\pm 0.07$  & $0.99\pm 0.08$ & $1.13\pm 0.08$ & $24\pm 9$ & $9\pm 9$ \\
\tableline \\[-1.0cm]
\end{tabular}
\tablecomments{All models are 2D to allow an independent comparison of
the data sets in the central regions. The iron abundance is expressed
in solar units. ``\%'' is the percent difference between the \xmm\ and
\chandra\ abundances. ``fix'' means that the relative normalizations
of the temperature components for the 2T models are required to have their
best-fitting values obtained from the joint \xmm\ and \chandra\ fits
to guarantee a consistent comparison.}
\end{center}
\end{table*}

\begin{figure*}[t]
\centerline{\psfig{figure=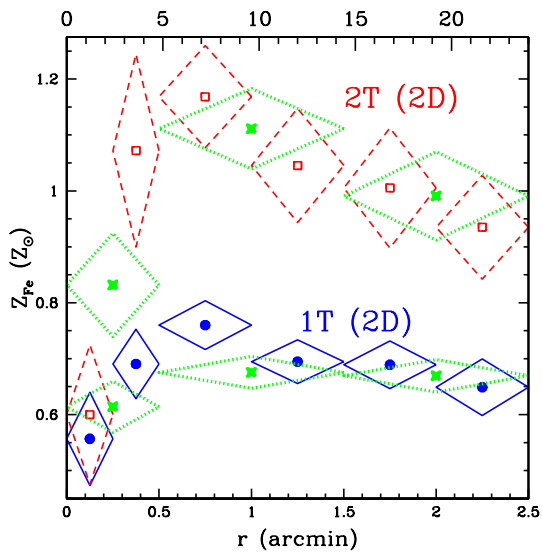,height=0.3\textheight}}
\caption{\label{fig.chan} Radial profiles
(units -- bottom: arcminutes, top: kpc) of the Fe abundance and
associated $1\sigma$ errors for 1T (blue circles and solid diamonds)
and 2T (red boxes and dashed diamonds) models. Only the \chandra\ data
are used and each model is 2D; i.e., not obtained from the spectral
deprojection analysis. We also show for each model (green crosses and
dashed diamonds) the results obtained in the wider apertures used in
the joint \xmm\ and \chandra\ fits; the results for these fits are
given in Table \ref{tab.chan}.}
\end{figure*} 

We have examined possible systematic errors in the measurements of the
metal abundances arising from calibration differences between the
\xmm\ MOS, \xmm\ pn, and \chandra\ ACIS-S3 CCDs. In Table
\ref{tab.chan} we list the values of \fe\ obtained from 1T (2D) and 2T
(2D) models fitted separately to the \xmm\ and \chandra\ data; i.e.,
the MOS and pn data were fitted simultaneously while the
\chandra\ data were fitted alone. We focus on 2D models so that the
fits for a specific annulus are independent of results obtained from
fits to adjacent regions at larger radii.

Using the 1T model the \xmm\ and \chandra\ data in annuli 1 and 3 give
values of \fe\ that agree within 5\%. In annulus 2 the value obtained
from \xmm\ is $17\%\pm 6\%$ lower than that obtained from
\chandra. The significance of this discrepancy is not very high.
However, it is noteworthy that in annulus 2 we find the largest
improvement for a 2T model over a 1T model (Paper 1). Given the very
poor fit of the 1T (2D) model within annulus 2 the modest discrepancy
of $17\%\pm 6\%$ between detectors could simply arise from the
different energy-dependent sensitivities of the detectors.

The effects of different sensitivities are illustrated by the 2T (2D)
fits also listed in Table \ref{tab.chan}. If the relative
normalizations of the two temperature components are allowed to be
different for \xmm\ and \chandra\ (i.e., denoted by ``free'' in Table
\ref{tab.chan}), then we find that the values of \fe\ obtained from
\xmm\ exceed those obtained by \chandra\ by 15\%-30\%. However, if we
fix the relative normalizations of the cooler and hotter components to
their best-fitting values obtained from the joint \xmm-\chandra\ fits
(i.e., denoted by ``fix'' in Table \ref{tab.chan}), then we find no
significant differences between the \xmm\ and \chandra\ fits; i.e.,
the percentage differences are all less than the (relatively large)
$1\sigma$ errors. Since the difference of the quality of the fits (not
shown) is quite negligible between the fixed and free cases, the
different values of \fe\ obtained in the two cases are probably
attributed to the different energy-dependent sensitivities of the
\xmm\ and \chandra\ detectors. We conclude that errors in the relative
calibration between \xmm\ and \chandra\ contribute errors in the
measured value of \fe\ of less than 20\%.

The agreement of \fe\ obtained from the \chandra\ and \xmm\ CCDs also
indicates that the larger PSF of \xmm\ does not significantly affect
our results; i.e., we have chosen bins that are sufficiently wide for
the \xmm\ PSF -- consistent with our findings regarding the
temperature profile in Paper 1. 

(We mention that we have performed an identical study to assess
differences in \fe\ measured between the \xmm\ MOS and pn CCDs.  For
annuli 5-8 we find that the values of \fe\ obtained from separate
fitting of the MOS and pn data generally agree within their 1-2
$\sigma$ errors. However, for annuli 1-4 the values of \fe\ obtained
from the pn data always exceed the values obtained from the MOS by as
little as 10\% to as much as 50\%, with the MOS values usually
agreeing better with the \chandra\ data in their regions of overlap.
As a result, we explored using only the MOS and \chandra\ data in our
analysis but found that the errors in the deprojected \fe\ abundances
were sufficiently large that we had to regularize the iron abundance
to obtain a radial profile similar to the 2D result. After this
regularization we noticed that the results were quite consistent with
the original un-regularized \fe\ values obtained from the simultaneous
MOS+pn fits. Because of this agreement, and the fact that the MOS+pn
fits do not require \fe\ to be regularized, the good agreement of the
MOS+pn fits with the \chandra\ fits discussed above, and the excellent
agreement of the MOS+pn and \asca\ data discussed in \S \ref{bias}, we
decided to use all of the data sets in our default analysis.)

As in Paper 1 we also performed fits to the \chandra\ data within
annuli that are half the size of those used for the \xmm\ and joint
\xmm\ -- \chandra\ fits. In Figure \ref{fig.chan} we
show \fe\ obtained from 1T (2D) and 2T (2D) fits to the \chandra\ data
in annuli of $30\arcsec$ width between $0.5\arcmin-2.5\arcmin$ and
$15\arcsec$ width between $0\arcsec-30\arcsec$ (see also Table 5 and
Figure 6 in Paper 1). Also shown in Figure \ref{fig.chan} are the
results using the larger $1\arcmin$-width annuli for the results
obtained from the joint \xmm\ -- \chandra\ fits. The values for \fe\
obtained in the larger bins are mostly consistent with an average
value of \fe\ obtained in the thinner bins. Moreover, if the hot gas
is actually single-phase, we would have expected in the smaller bins
to have a smaller radial temperature variation, and thus a weaker ``Fe
Bias'', and therefore a smaller underestimate of \fe\ using the 1T
model. Instead, we see good agreement between the \fe\ values obtained
using the thinner and wider bins. (Note: consistent results are
obtained with 3D models using the \xmm\ data to deproject shells
exterior to $r=2.5\arcmin$.) This agreement is consistent with the
models for limited multiphase gas (i.e., particularly the 2T and
PLDEM models.)

In Figure \ref{fig.chan} one also notices that the central dip in \fe\
is apparently a smooth transition beginning within the \chandra\
annulus $R=15\arcsec-30\arcsec$ and finishing within the central
bin. It is within the central bin ($R=15\arcsec$) where the value of
\fe\ obtained from the 1T and 2T models (2D) agree best.

As discussed in \S 6.1.2 of Paper 1 the value of \fe\ obtained from
the \xmm\ RGS data of NGC 5044 by \citet{tamu03a} is in excellent
agreement with the value we obtain from the simultaneous fits to the
\xmm\ and \chandra\ CCDs in the central radial bin. 
We emphasize that this excellent agreement occurs when the same models
(with the same free parameters) are fitted over the same energy range
for both the RGS and the CCDs. The good agreement provides a further
check on possible calibration differences between detectors.

We conclude that both the \xmm\ and \chandra\ CCD data require \fe\ to
be at least solar within $r\sim 30$~kpc except for within the very
central region ($r=2.4$~kpc).

\subsection{Plasma Codes}
\label{plasma}

\begin{figure*}[t]
\parbox{0.49\textwidth}{
\centerline{\psfig{figure=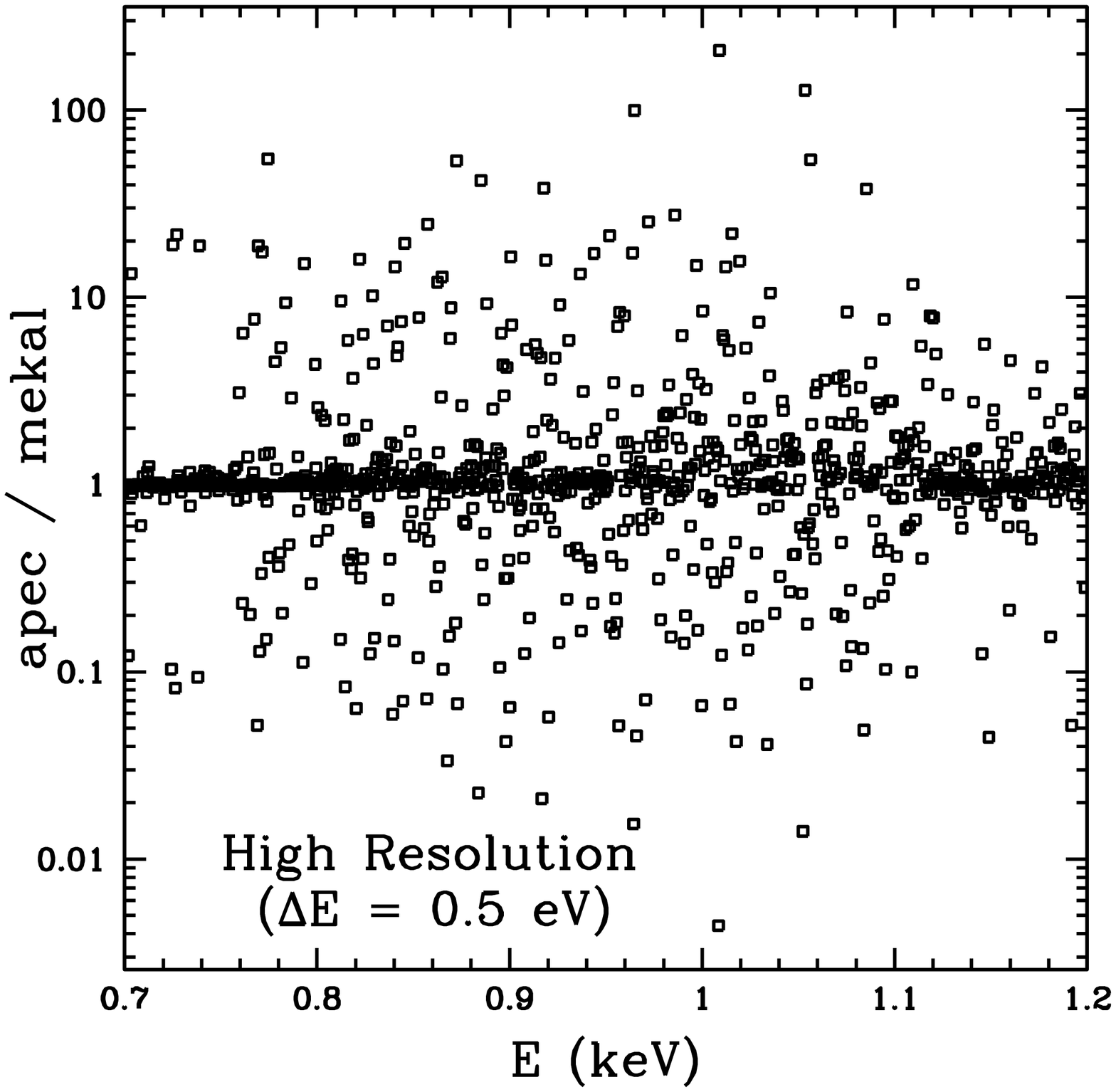,height=0.3\textheight}}}
\parbox{0.49\textwidth}{
\centerline{\psfig{figure=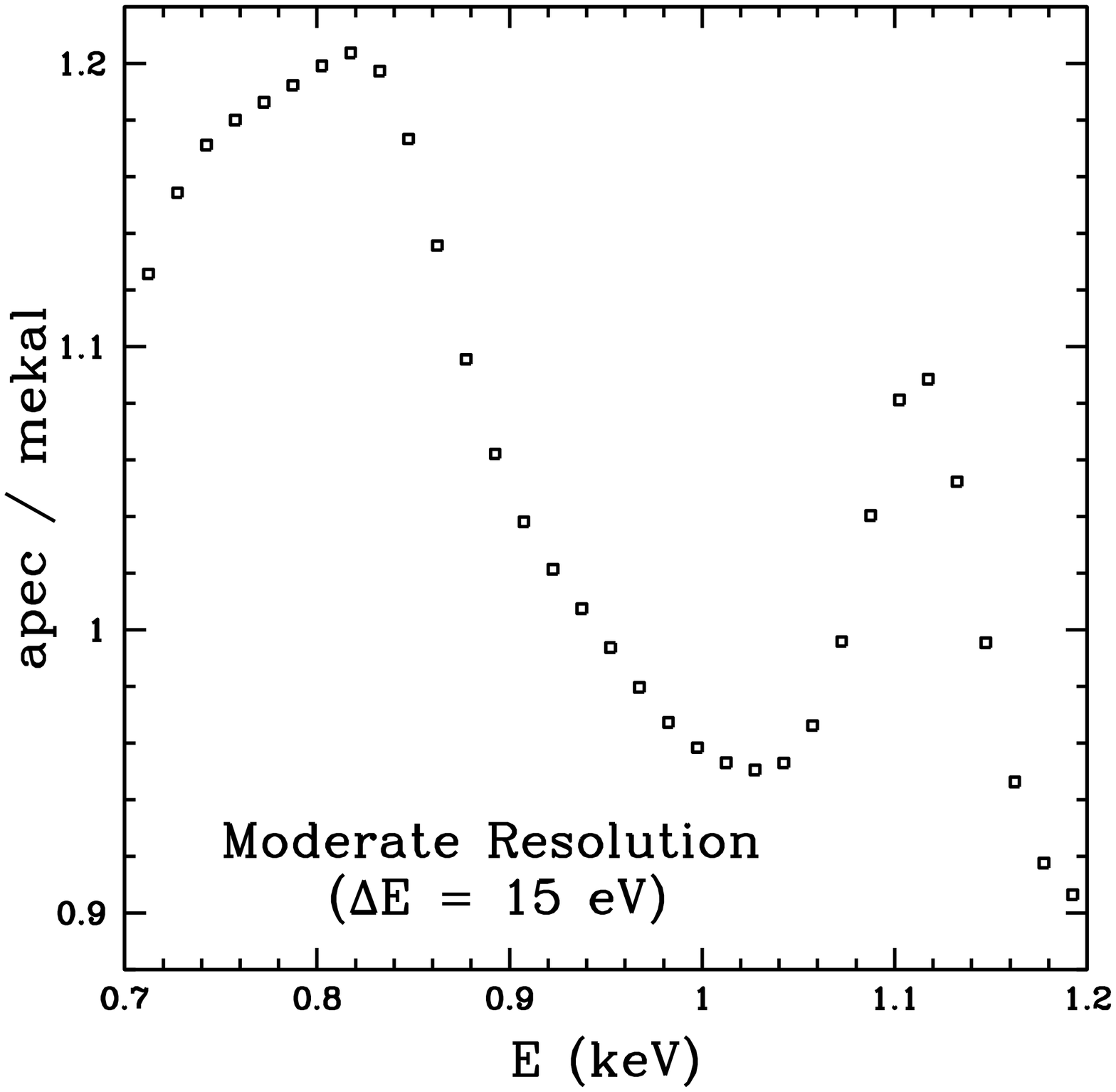,height=0.3\textheight}}}

\caption{\label{fig.fel} Comparison of the \apec\ and \mekal\ plasma
codes in the energy region, 0.7-1.2~keV, where the Fe L shell emission
lines dominate for a $\sim 1$~keV plasma. We show the \apec/\mekal\
ratio for a $T=1$~keV, solar metallicity plasma ({\sl Left panel}) at
high resolution and ({\sl Right panel}) at moderate resolution, the latter
corresponding approximately to the response matrix binning of the MOS
data of NGC 5044 we used in our spectral fitting.}
\end{figure*} 

For a single isolated emission line of a particular element in a
coronal plasma the abundance of that element may be directly measured
by calculating the ratio of the flux within the line to the flux of
the local continuum. In such an idealized case, any error in the
theoretical calculation of the line flux within the plasma code will
translate directly to an error in the measured elemental abundance.
Consider, however, the case of the Fe L shell lines for a system like
NGC 5044. 

In a coronal plasma with $T\sim 1$~keV there is a forest of Fe L lines
near $1$~keV, with the most important lines spanning the approximate
range 0.7-1.2~keV. Consider also these lines when observed at the
moderate spectral resolution of the \xmm\ and \chandra\ CCDs ($\Delta
E \approx 50-100$~eV near 1~keV). In these cases the Fe abundance is
obtained by fitting data over a broad energy range, so that all of the
strong Fe L lines over 0.7-1.2~keV play a key role as does the
continuum determined from regions outside the Fe L region. The net
effect is that even rather large errors in a small number of lines do
not necessarily translate directly to large errors in the inferred Fe
abundance.

We illustrate this effect in Figure \ref{fig.fel} using the \apec\ and
\mekal\ plasma codes.  These codes have many differences in
how they model the atomic physics and in their line libraries
(especially for Fe L) thus allowing for an interesting test of the
robustness of the iron abundance determination.  We consider a
fiducial spectral model with $T=1$~keV, $Z=1\solar$, and equal
normalizations for the \apec\ and \mekal\ codes.  The ratio of the
fluxes in the two models are plotted as a function of energy over the
range 0.7-1.2~keV for two different resolutions. In the left panel we
plot this ratio at high-resolution binning ($\Delta E=0.5$~eV) giving
1000 energy bins over the energy range shown. Although most of the
points cluster near 1, there are many points having large errors;
i.e., ratios larger than 2 or less than 1/2. In the right panel we
plot the same ratio at the binning for the moderate resolution CCDs
($\Delta E=15$~eV) used in our analysis of NGC 5044; specifically,
these energy bin sizes correspond to those of the re-binned MOS1 data
in the central radial bin for NGC 5044. In this case the ratios are
all within $\approx 20\%$, and thus we should expect that the iron
abundances should not differ by much more than this amount for each
model.

Indeed, we find that both 1T and 2T models (each 2D) computed using
the \apec\ code differ by no more than 10\%-20\% from those computed
using the \mekal\ code. Interestingly, the small differences between
\apec\ and \mekal\ do appear to be a real systematic effect despite
the fact that these differences are consistent within the $1\sigma$
errors. The sign of the systematic difference is opposite for 1T and
2T models: for 1T models, $\fe^{\rm apec}> \fe^{\rm mekal}$, while for
2T models, $\fe^{\rm apec}<
\fe^{\rm mekal}$.

We conclude that the measured iron abundances are accurate to within
10\%-20\% considering reasonable remaining errors in the plasma codes
such as those associated with the \ion{Fe}{17} lines near 0.7~keV
discussed by \citet{beha01}. Of course, for high-resolution studies
using the gratings on \chandra\ and \xmm\ more care must be taken when
analyzing the properties of individual lines such as done by
\citet{beha01}. But even high-resolution studies of NGC 5044 using the
RGS obtain fully consistent values for \fe\ from broad-band spectral
fitting with the values obtained from the \xmm\ and \chandra\ CCDs as
discussed at the end of \S \ref{calib}.

\subsection{Bandwidth}
\label{emin}

\begin{figure*}[t]
\centerline{\psfig{figure=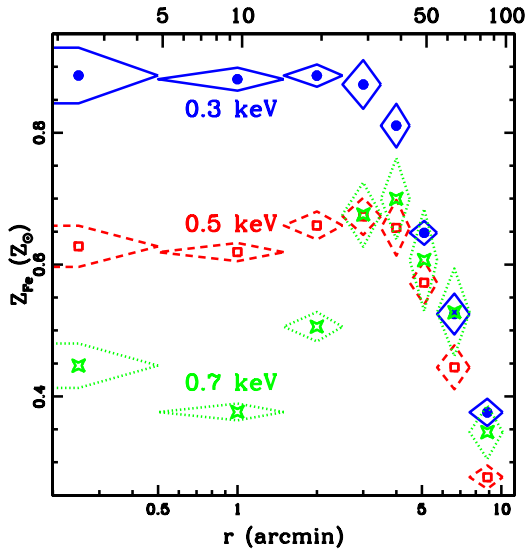,height=0.3\textheight}}
\caption{\label{fig.emin} Radial profiles (units -- bottom:
arcminutes, top: kpc) of the Fe abundance and 
associated $1\sigma$ errors for the 1T (2D) model obtained from
fitting data with lower energy limit $\emin=0.3,0.5,0.7$~keV. }
\end{figure*} 

In section 5.1 of \citet{buot00c} we discussed the sensitivity of the
Fe Bias to the lower energy limit used in the spectral fitting of a
$T\sim 1$~keV coronal plasma. Here we summarize this discussion and
refer the interested reader to the previous paper for
details. Consider (1) a 1T model ($T=1$~keV) of a coronal plasma with
solar metallicity, and (2) a multitemperature coronal model with
temperature components distributed across $T\approx 0.7-1.5$~keV also
with solar metallicity; e.g., the simplest case of a 2T model with
$\tcool=0.7$~keV and $\thot=1.5$~keV each with the same emission
measure. Each model has the same total emission measure. The spectral
energy distribution of the 1T model has a much narrower peak near
1~keV than the 2T model; see Figure 5 of
\citet{buot00c}.

If the observed spectrum is similar to the 2T model, such as we
observe for NGC 5044 with \xmm\ and \chandra\ within $r\sim 30$~kpc,
then the 1T model with solar metallicity is a poor fit to the data. If
the iron abundance is allowed to vary, then the fitting software
(\xspec\ in this case) tries to improve $\chi^2$ near 1~keV by
reducing the value of \fe\ while simultaneously raising the continuum
to maintain the quality of the fit near the wings of the Fe L
region. The resulting Fe abundance obtained with the 1T model is
therefore an underestimate of the true value. It is this effect that
we have previously termed the ``Fe Bias''
\citep[e.g.,][]{buot00a,buot00c}.

However, the ability of the $\chi^2$ fitter to increase the
contribution of the continuum within the Fe L region depends
sensitively on the constraints on the continuum outside the Fe L
region. At higher energies the model compensates for this increase in
the continuum by simultaneously decreasing the abundances of Mg, Si,
and S while maintaining an approximately fixed $\alpha$/Fe ratio as
discussed in \S \ref{si} and \ref{other}; note that Mg, Si, and S have
the strongest emission lines at energies above the Fe L region in NGC
5044. 

At energies below the Fe L region (i.e., below 0.7~keV), the strongest
line is far and away \ion{O}{8} ly-$\alpha$ near 0.65~keV -- very
close to the bottom edge of the Fe L region. Below the \ion{O}{8} line
down to the $\sim 0.3$~keV bandpass limits of the \xmm\ and \chandra\
CCDs there are no strong emission lines. (Allowing the C and N
abundances to vary affects the inferred value of \fe\ by $<10\%$.)
Therefore, possible changes in the continuum level below the
\ion{O}{8} ly-$\alpha$ line requested by the $\chi^2$ fitter to better
match the Fe L region cannot be compensated for by changes in the
elemental abundances. Changing the temperature is also not an option
since it is tightly constrained by the shape of the continuum both
above and below the Fe L region. (It is found that the inferred
temperature of the 1T model is quite insensitive to the fitted -- or
assumed -- value of \fe\ over a large range in \fe.)  Hence, below
$\approx 0.7$~keV the continuum data are a strong inhibitor of the Fe
Bias, though they do not prevent it entirely as demonstrated by the
differences we obtained for 1T and multitemperature models discussed
in \S \ref{fe}.

Consequently, it is expected that for spectra where the Fe Bias is
unimportant (e.g., the \xmm\ spectra for $r\ga 40$~kpc in NGC 5044)
the measured value of \fe\ should not be very sensitive to the value of
\emin\ -- the lower energy limit (0.3-0.7~keV) used in the spectral
fitting. In contrast, for spectra where the Fe Bias is important
(e.g., the \xmm\ spectra for $r\la 30$~kpc in NGC 5044) the measured
value of \fe\ should be very sensitive to the value of \emin, and
specifically \fe\ should decrease as \emin\ increases toward 0.7~keV.

This behavior is observed for NGC 5044. In Figure \ref{fig.emin} we
plot $\fe(R)$ for the 1T (2D) model fitted jointly to the \chandra\
and \xmm\ data for different values of $\emin=0.3,0.5,0.7$~keV. For
$R\ga 40$~kpc the values of \fe\ for $\emin=0.3,0.7$~keV agree within
their $\sim 1\sigma$ errors. The values for $\emin=0.5$~keV are
somewhat below these values because of the degeneracy with the oxygen
abundance. (If the oxygen abundance is tied to the iron abundance in
their solar ratio then the \fe\ values for $\emin=0.5$~keV are within
10\%-20\% of the values of the other \emin.) For $R\la 30$~kpc, where
we have the evidence for multitemperature gas and the corresponding Fe
Bias, we observe that \fe\ decreases as \emin\ increases toward
0.7~keV as expected.

We mention that the \fe\ values inferred from 2T models also follow
the same trends, though not as dramatically as for the 1T models
within $R\la 30$~kpc.  The reason why 2T models are affected at all is
that the need for multitemperature models decreases (though is not
removed) as \emin\ approaches 0.7~keV as discussed in Section 4.4 of
Paper 1. Since a single temperature component dominates more in the
fits for larger \emin, the multitemperature models suffer more from
the Fe Bias for larger \emin. For example, the best-fitting values of
\fe\ for the 2T (2D) model in radial bin \#2 are respectively
$(1.38,1.19,0.86)$ in solar units for $\emin=0.3,0.5,0.7$~keV. These
values still exceed their respective purely 1T (2D) counterparts by a
factor of $\approx 2$.

Hence, for $T\sim 1$~keV plasmas consisting of multiple temperature
components distributed across $T\approx 0.7-1.5$~keV, we conclude that
reliable measurements of \fe\ can only be obtained for \emin\ below
$\approx 0.5$~keV since the continuum emission below 0.6~keV serves as
an important check on the Fe Bias. In this paper we have decided to
emphasize results obtained for $\emin=0.5$~keV rather than for
$\emin=0.3$~keV because of the somewhat better agreement obtained for
the values of \fe\ between the \xmm\ and \chandra\ detectors for
$\emin=0.5$~keV discussed in \S \ref{calib}. However, the 10\%-20\%
larger values of \fe\ obtained from the joint \xmm-\chandra\ fits for
$\emin=0.3$~keV should be considered a reasonable estimate of the
systematic error arising from the choice of \emin\ for the results
quoted in this paper.

\subsection{Variable \nh\ and Intrinsic Absorption}
\label{nh}

Allowing for absorption by cold gas ($T\la 10^4$~K) with a hydrogen
column density (\nh) in excess of the Galactic value ($\nhgal=5\times
10^{20}$~\cmsq) affects the values of \fe\ obtained from
single-temperature models in much the same way as increasing \emin\
toward 0.7~keV with $\nh=\nhgal$ as discussed above in \S
\ref{emin}. That is, the values of \fe\ obtained from 1T fits with
$\emin=0.7$~keV and $\nh=\nhgal$ are broadly similar to those obtained
from 1T fits with \nh\ in excess of \nhgal\ for $\emin=0.3-0.7$~keV;
our fiducial absorber model is a foreground screen ($A(E) =
\exp(-\nh\sigma(E))$) with separate values of \nh\ for each annulus,
although we have explored a suite of absorber models (e.g., with
redshift at NGC 5044) and have obtained fully consistent results.

As discussed in \S 6.4 of Paper 1 the 2T model is still clearly
preferred over the 1T model when allowing for variable \nh\ in each
case. Unlike for the 1T models, the values of \fe\ obtained for the
multitemperature models with variable \nh\ only differ significantly
with the corresponding model with Galactic \nh\ in shell \#2; e.g.,
for the 2T (2D) model we have, $\fe = 0.96\pm 0.05\solar$ (variable
\nh) and $\fe=1.19\pm 0.06\solar$ (Galactic \nh) -- though in both
cases \fe\ is near the solar value. Since (as discussed in Paper 1)
the multitemperature models with $\nh=\nhgal$ provide better fits than
1T models with variable \nh\ within the central $\approx 30$~kpc,
there is no obvious sharp absorption feature in the spectrum, and
there is no evidence from observations in other wavebands for the
large quantities of cold absorbing material ($>10^{9}\msun$) implied
by the fitted values of \nh, we do not take seriously the results
obtained from the intrinsic cold absorber models.

We mention that a collisionally ionized ``warm'' absorber ($T\approx
10^{5-6}$~K) model, in contrast to the cold absorber, does not affect
significantly energies below $\sim 0.5$~keV and results in fitted
values of \fe\ quite similar to the models without any intrinsic
absorption.  Another interesting feature of the warm absorber is that
the fitted oxygen abundances in the hot gas in the central regions are
$\approx 0.6\solar$ and the Mg/O ratios are near solar. The near-solar
Mg/O ratios are in much better agreement with expectations from SNe
enrichment than the values obtained for all other models we explored
without a warm absorber.  However, like the cold absorber there is no
evidence for the emission in other wavebands implied by the relatively
large column densities obtained in the X-ray fits for the warm
absorber, nor is it easy to understand how the temperature of the warm
gas is maintained. Consequently, we do not discuss either the warm or
cold intrinsic absorbers further.

\subsection{Background}

We considered the effect of errors in the background normalization on
the measured values of \fe. Apart from the few emission lines from
calibration sources in the \xmm\ and \chandra\ CCDs, the background is
a smooth and slowly varying function of energy over the 0.3-5~keV
bandpass. Consequently, if the background contribution to the spectrum
is underestimated, one will believe the continuum emission is larger
than in reality. As a result, one will mistakenly infer smaller
equivalent widths for the emission lines and therefore underestimate
the values for the metal abundances. Over-subtraction of the
background leads to larger equivalent widths and overestimates of the
metal abundances.

To examine the sensitivity of the measured values of \fe\ to
reasonable background errors we repeated our analysis using the
background templates renormalized to have count rates +20\% and -20\%
of their nominal values. By fitting 1T (2D) and 2T (2D) models to the
background-subtracted data using these renormalized templates, we
calculated the variation in \fe\ as a function of radius for each
case. In the outer annuli (bins \#7 and \#8), where the background is
most important, we find that the measured values of \fe\ change by
$<\pm 15\%$ with respect to the nominal background case. The variation
is progressively smaller at smaller radii being $<\pm 1\%$ in the
central bin.

We mention that even extreme errors in the background do not generate
qualitatively different results in the measured values of \fe,
especially within the central $R\sim 30$~kpc. For example, if we do
not subtract the background at all the value of \fe\ is underestimated
by $<10\%$ for 1T (2D) models and overestimated by $<5\%$ for 2T (2D)
models within radial bins 1-3. The reason why the 2T models here give
a slight overestimated of \fe\ when the background is under-subtracted
is because the excess continuum at higher energies is interpreted as
part of the higher temperature component. The increased normalization
of this component causes an increase in the inferred value of \fe\
according to the Fe Bias mechanism (e.g., see discussion at the
beginning of \S \ref{emin}). At the largest radius (bin 8), if the
background is not subtracted at all we obtain best-fitting values:
$\fe=0.17\solar$ compared to the nominal value of $\fe=0.28\solar$ for
1T, and $\fe=0.33\solar$ compared to the nominal value of
$\fe=0.36\solar$ for 2T.

\section{Emission-Weighted Average Abundances and The Fe Bias Revisited}
\label{bias}

\subsection{Central Regions}

\begin{figure*}[t]
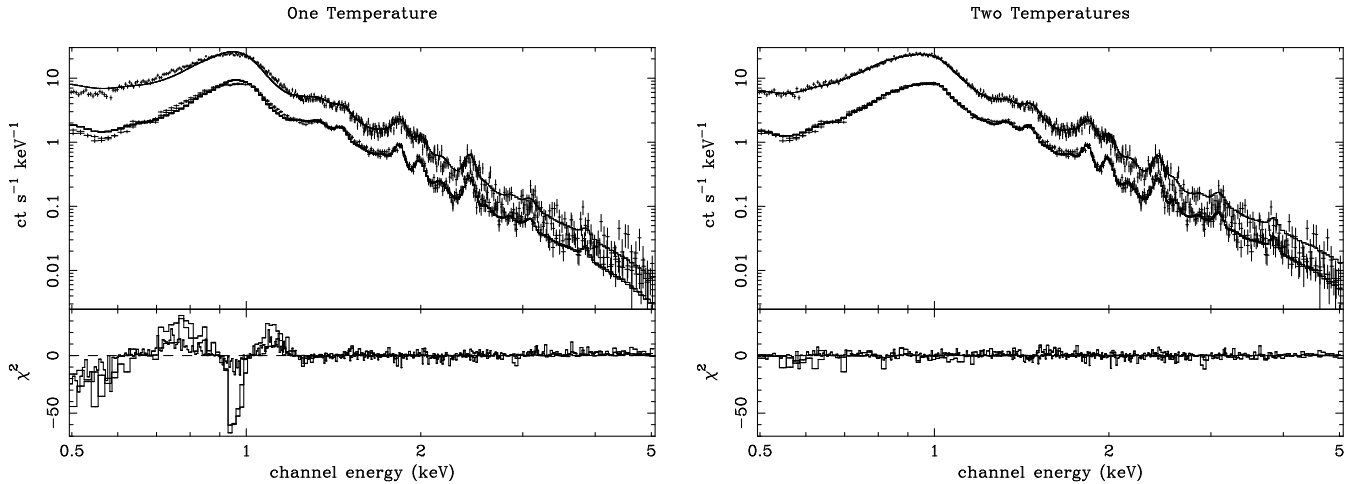

\parbox{0.49\textwidth}{
\centerline{\psfig{figure=f8a.eps,angle=-90,height=0.26\textheight}}}
\parbox{0.49\textwidth}{
\centerline{\psfig{figure=f8b.eps,angle=-90,height=0.26\textheight}
}}
\caption{\label{fig.febias} MOS1, MOS2, and pn spectra accumulated
within a circular aperture of radius, $R=5\arcmin$ (48~kpc), fitted
with ({\sl Left panel}) a single temperature (1T) model and ({\sl
Right panel}) a two-temperature (2T) model; both models are 2D. In
each case the \apec\ plasma model is used and the solar abundances are
taken from \citet{grsa} which use the new (smaller) photospheric value
for the iron abundance. The models are the same as discussed in \S
\ref{fe} for the spatially resolved analysis. That is, Galactic
absorption is assumed and the following metal abundances are free
parameters: O, Ne, Mg, Si, S, Fe and all other abundances are tied to
Fe in their solar ratios. For the 2T model the abundances of each
temperature component are tied together in the fits.}
\end{figure*}

\begin{table*}[t] 
\caption{Emission-Weighted Average Abundances Within $R=5\arcmin=48$~kpc
\label{tab.abun2}} 
\begin{center} \vskip -0.4cm
\begin{tabular}{cccccccc} \tableline\tableline\\[-7pt]
& Best & $\Delta$Statistical & $\Delta$Model & $\Delta$Plasma Code & $\Delta$\emin\ & $\Delta\nh$ & $\Delta$Background
\\
\tableline \\[-7pt]
\fe\ & 1.09 & $\pm 0.04$ & $\pm 0.01$ & $-0.05$ & $+0.23$         & $\pm 0.03$  & $\pm 0.01$ \\
\si/\fe & 0.832 & $\pm 0.017$ & $\pm 0.004$ & $+0.058$ & $+0.013$ & $\pm 0.011$ & $\pm 0.003$\\
\su/\fe & 0.542 & $\pm 0.020$ & $\pm 0.006$ & $+0.028$ & $-0.021$ & $\pm 0.002$ & $\pm 0.002$\\
\ox/\fe & 0.331 & $\pm 0.028$ & $\pm 0.005$ & $-0.052$ & $+0.089$ & $\pm 0.014$ & $\pm 0.003$\\
\mg/\fe & 0.874 & $\pm 0.026$ & $\pm 0.013$ & $+0.026$ & $+0.055$ & $\pm 0.023$ & $\pm 0.005$\\
$Z_{\rm Ne}$ & 0.56 & $\pm 0.09$ & $\pm 0.10$ & $-0.37$ & $+0.26$ & $\pm 0.04$  & $\pm 0.03$ \\

\tableline \\[-1.0cm]
\end{tabular}
\tablecomments{Selected abundances and ratios obtained from fitting
the accumulated \xmm\ EPIC MOS1, MOS2, and pn spectral data within a
circular aperture of radius $5\arcmin$ (48~kpc). The ``Best'' column
indicates the average of the best-fitting values obtained from the 2T
and PLDEM \apec\ models expressed in terms of the solar abundance
table of \citet{grsa}. The ``Statistical'' column gives the
statistical error. ``Model'' is the difference in the best-fitting
values obtained from the 2T and PLDEM models. ``Plasma Code''
represents the difference obtained when using the \mekal\ plasma
code. ``\emin'' is the difference when setting the lower energy limit
to 0.3~keV. ``$\nh$'' indicates the error if the Galactic absorption
column density is set to $\pm 10\%$ of the assumed value of $5\times
10^{20}$~\cmsq. ``Background'' represents the error if the background
level is set to $\pm 20\%$ of nominal. Calibration error should
contribute less than a 10\% error on \fe\ considering comparisons to
\asca\ (\S \ref{bias}) and \chandra\ data (\S \ref{calib}) of NGC 5044.}
\end{center}
\end{table*}

Irrespective of whether the hot gas at each radius is actually
single-phase or multiphase, all models of the spatially varying
spectra that we investigated in Paper 1 require that the accumulated
spectra within $R\sim 30$~kpc contain a range of temperature
components with $T\approx 0.7-1.2$~keV. Fitting a 1T model to this
entire accumulated spectrum must fail according to the Fe Bias as
summarized at the beginning of \S \ref{emin}.

We demonstrate this manifestation of the Fe Bias as follows. We
extracted the accumulated spectra of the \xmm\ EPIC MOS1, MOS2, and pn
data within a radius of $5\arcmin$ (48~kpc); we choose a radius
slightly larger than 30~kpc to fully enclose the peak of the 1T
temperature profile (Paper 1, Figure 3) and to facilitate comparison
to previous \asca\ results below. (Note the \chandra\ ACIS-S3 data do
not extend to this radius.) The result of fitting a 1T \apec\ model
simultaneously to the MOS1, MOS2, and pn data is shown in Figure
\ref{fig.febias}.  Readily apparent are the residuals characteristic
of the Fe Bias seen in the smaller apertures in Paper 1 and in the
larger apertures in our previous \asca\ studies of the brightest
elliptical galaxies in centrally E-dominated groups (see especially
Figure 5 in \citealt{buot00c} and the Appendix in \citealt{buot00a}).
The fit is formally of very poor quality ($\chi^2=3354.5$ for 816 dof)
and the iron abundance is $0.51\pm 0.01\solar$. We reiterate that the
poor fit and low value of \fe\ are exactly as expected because of the
presence of multitemperature temperature components\footnote{We note
that allowing \nh\ to vary barely improves the fit: $\chi^2=2958.1$
for 815 dof. The fitted absorption column density is $\Delta\nh\approx
5\times 10^{20}$~\cmsq\ above the Galactic value, implying large
amounts of cold gas that have never been observed in other wavebands
in NGC 5044 or similar systems; i.e., $M_{\rm abs}\approx m_{\rm
H}\Delta\nh\pi (48\, \rm kpc)^2 \approx 3 \times 10^{10}\msun$
assuming solar abundances and that the absorber is uniformly
distributed. If the abundance ratios are similar to the hot gas, then
$\ox/\fe\approx 0.2$ implies $M_{\rm abs}\approx 10^{11}\msun$ because
oxygen is the primary absorber. If the absorber is not uniformly
distributed then the quoted values for $M_{\rm abs}$ are only lower
limits.} within the aperture -- {\bf but we have made no assumption
about whether these components arise from a radially varying
single-phase or a true multiphase medium.}

A vastly improved fit which (1) eliminates the residuals near 1 keV to
a magnitude similar to that present at other energies, and (2)
provides a value of \fe\ over twice as large as the 1T value is
obtained using a simple 2T model (i.e., discrete temperature
distribution) or a PLDEM (i.e., continuous temperature
distribution). Each of these models adds only 2 free parameters over
those of the 1T model. In Figure \ref{fig.febias} we display the 2T
model fitted to the total accumulated EPIC spectra within 48~kpc.  It
is clear that the residuals near 1~keV are greatly reduced the fit is
improved dramatically ($\chi^2=1202.3$ for 814 dof). The temperatures
obtained for the 2T model are $\tcool=0.804\pm 0.004$~keV and
$\thot=1.38\pm 0.02$~keV which are very similar to the range of values
obtained from the spatially resolved 1T and 2T models (Paper 1); note
that the best-fitting ratio of emission measures of the hotter and
cooler components is 1.04. Moreover, the iron abundance for the 2T
model is a factor of 2.1 times larger than obtained for the 1T model:
$\fe=1.08\pm 0.04\solar$ (statistical error). The PLDEM gives a fit
and abundance values extremely similar to the 2T model:
$\chi^2=1235.1$ for 814 dof and $\fe=1.10\pm 0.04\solar$.  (Note we
obtain the following temperature parameters for the PLDEM: $\alpha=
-1.2\pm 0.2$, $\tmin = 0.665\pm 0.008$~keV, $\tmax-\tmin= 1.12\pm
0.05$~keV.)

The superb agreement between the 2T and PLDEM models demonstrates that
the EPIC spectral data accumulated within 48~kpc have both sufficient
sensitivity and resolution to unequivocally rule out the 1T model (as
expected) while constraining the DEM well enough so that fully
consistent values for the abundances are obtained using very different
multitemperature models; i.e., the average (emission-weighted)
abundances within 50~kpc are very well constrained by the EPIC data.

In Table \ref{tab.abun2} we list the value of \fe\ obtained from the
multitemperature fits within 50~kpc and present a detailed accounting
of the error budget following our discussion in \S \ref{sys}. It is
noteworthy that the largest source of error is \emin\ which leads to a
larger value.  The emission-weighted value of \fe\ is a consistent
average of the values within 50~kpc obtained from the spatially
resolved analysis (Figure
\ref{fig.fe}).

Also shown in Table \ref{tab.abun2} are the emission-weighted average
abundance ratios \si/\fe, \su/\fe, \ox/\fe, and \mg/\fe\ obtained
within 50~kpc and their associated error budgets. These ratios are
very tightly constrained and agree with the mass-averaged values
obtained from the spatially resolved analysis (\S \ref{si} and
\ref{other}). We mention that the \mg/\fe\ ratio is well constrained,
and the only source of error we believe could possibly account for its
relatively large value is incomplete subtraction of the Al calibration
lines near 1.4~keV. However, the small error in the abundance ratios
associated with reasonable background errors make this explanation
seem implausible as well.

The Ne abundance and its error budget are also shown in Table
\ref{tab.abun2}. Since the key Ne emission line is well hidden within
the Fe L forest, the value of \neon\ is quite susceptible to
differences in the temperature model and plasma code. In fact, it
exhibits by far the largest differences between the \apec\ and \mekal\
plasma code. Despite the relatively large systematic errors, the value
of \neon\ is clearly sub-solar with probably values between 0.3-0.5
solar. 

Although we have established that calibration errors should contribute
at most 20\% extra uncertainty in the measured value of \fe\ (\S
\ref{calib}), we can provide a further calibration check by comparing
to our previous \asca\ studies of NGC 5044 \citep{buot98c,buot99a}.
\citet{buot98c} analyzed the \asca\ SIS0 and SIS1 spectra accumulated
within circles of $R \approx 5\arcmin$ centered on NGC 5044. They
fitted \mekal\ plasma models over 0.5-5~keV where (1) all abundances
were tied to iron in their solar ratio, (2) the solar abundance table
of \citet{angr} was used, and (3) the absorption column density was a
free parameter. They obtained $\fe=0.25\solar$ for the 1T model (no
error given because of the poor fit) and
$\fe=0.62^{+0.11}_{-0.08}\solar$ (90\% confidence) for the 2T model;
the 2T model was a clearly superior fit (see Figures 1 and 5 of
\citealt{buot98c}). If we perform fits to the
\xmm\ EPIC data in the same region using exactly the same models we obtain
$\fe=0.256\pm 0.003\solar$ for 1T and $\fe=0.64\pm 0.02\solar$ for 2T
(only statistical error quoted) in excellent agreement with our
previous results from \asca. (Similar agreement is obtained when
comparing identical models to \citealt{buot99a}.) {\bf The excellent
agreement between \xmm\ and \asca\ implies that calibration error
cannot be a large contributor to error in our measurement of \fe,
certainly $<10\%$.}

Hence, the \xmm\ data of NGC 5044, whether extracted in a large 50~kpc
aperture as done in this section or in the smaller apertures in
previous sections, fully confirm the Fe Bias effect, not only for NGC
5044, but by implication also for many other bright ellipticals in
centrally E-dominated groups we examined in previous \asca\ studies
\citep{buot98c,buot99a,buot00a}.

Our results in this section clearly rule out the claim by
\citet{loew02} and \citet{mush03a} that the hot gas in the central
regions of very luminous galaxies like NGC 5044 must be spatially
isothermal. (Our spatially resolved analysis of the temperature
distribution in Paper 1 indicates that the gas is not single-phase as
well.) Previous \rosat\ and \asca\ studies that attempted to measure
\fe\ using 1T models necessarily obtained values that were biased low
\citep[e.g.,][]{awak94,mats94,fuka96,mats97,arim97,loew02,davi99,fino99}.
(We note that \citealt{mats00} obtained near-solar values for \fe\ in
some bright galaxies with \asca\ by adding a uniform 20\% systematic
error across the Fe L region. They assumed this systematic error arose
from calibration error or errors in the plasma code rather than from
multiple temperature components within a large aperture.)

We mention that although the 2T and PLDEM models provide vast
improvements in the fits over the 1T model, their $\chi^2$/dof values
are still formally unacceptable. The residuals of these fits do not
show any obvious features and have the same magnitude throughout the
bandpass. The ratios of model-to-data are within $\pm 5\%$ and appear
to be largely the result of small inconsistencies between the MOS1,
MOS2, pn, and ACIS-S detectors. (Formally acceptable $\chi^2$ values
are obtained, e.g., using just the \chandra\ data as shown in Table 4
of Paper 1.) However, since there are radial variations of the
spectral properties within the large aperture, we expect that the
models fitted to the entire large aperture must fail at some level;
i.e., the temperature components of the 2T model shown in Table
\ref{tab.temps} do vary with radius within $R=5\arcmin$.

\subsection{Outer Regions}

\begin{figure*}[t]
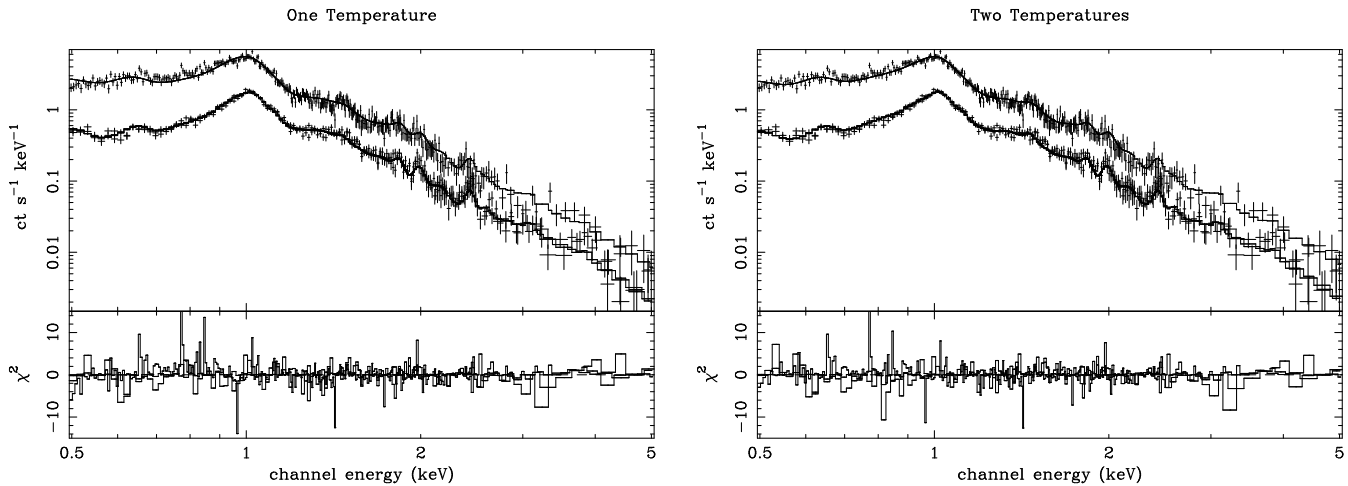

\parbox{0.49\textwidth}{
\centerline{\psfig{figure=f9a.eps,angle=-90,height=0.26\textheight}}}
\parbox{0.49\textwidth}{
\centerline{\psfig{figure=f9b.eps,angle=-90,height=0.26\textheight}
}}
\caption{\label{fig.febiasbig} Same as Figure \ref{fig.febias} except
the MOS1, MOS2, and pn spectra are accumulated
within an annulus, $R=5\arcmin-10\arcmin$ (48~kpc - 96~kpc). Notice
the Fe L lines are more peaked near 1~keV than in Figure
\ref{fig.febias} implying a narrower range of temperatures. The
smaller size of the ``hump'' of Fe L lines with respect to the
continuum also implies a lower value for the Fe abundance in this
case.}
\end{figure*}

\begin{table*}[t] 
\caption{Emission-Weighted Average Abundances Within $R=5\arcmin -
10\arcmin$ (48~kpc - 96~kpc)
\label{tab.abun3}} 
\begin{center} \vskip -0.4cm
\begin{tabular}{cccccccc} \tableline\tableline\\[-7pt]
& Best & $\Delta$Statistical & $\Delta$Model & $\Delta$Plasma Code & $\Delta$\emin\ & $\Delta\nh$ & $\Delta$Background
\\
\tableline \\[-7pt]
\fe\ & 0.44 & $\pm 0.02$ & $\pm 0.05$ & $+0.04$ & $+0.09$         & $\pm 0.01$  & $\pm 0.04$ \\
\si/\fe & 0.732 & $\pm 0.066$ & $\pm 0.020$ & $+0.003$ & $+0.028$ & $\pm 0.015$ & $\pm 0.002$\\
\su/\fe & 0.55 & $\pm 0.09$ & $\pm 0.01$ & $+0.01$ & $-0.02$      & $\pm 0.003$  & $\pm 0.12$ \\
\ox/\fe & 0.96 & $\pm 0.11$ & $\pm 0.02$ & $-0.11$ & $+0.18$      & $\pm 0.04$  & $\pm 0.10$ \\
\mg/\fe & 0.56 & $\pm 0.11$ & $\pm 0.12$ & $-0.16$ & $+0.15$      & $\pm 0.04$  & $\pm 0.20$ \\
$Z_{\rm Ne}$ & 0.20 & $\pm 0.11$ & $\pm 0.18$ & $-0.11$ & $+0.24$ & $\pm 0.04$  & $\pm 0.17$ \\
\tableline \\[-1.0cm]
\end{tabular}
\tablecomments{The columns are the same as described in in Table
\ref{tab.abun2} with one exception: ``Model'' refers to the average
value of the 1T, 2T, and PLDEM \apec\ models. This average
is calculated by first computing the average for the 2T and power-law
DEM to get the average multitemperature value. Then this value
is averaged with the 1T value to obtain the final average
value. Hence, ``Model'' mostly reflects the difference between 1T and
the multitemperature models.} 
\end{center}
\end{table*}

To obtain the average properties at large radius we extracted the EPIC
spectra within an annulus of $R=5\arcmin-10\arcmin$ (48~kpc - 96~kpc)
and performed the same analysis just described above for the central
region. The results of fitting 1T and 2T \apec\ models are shown in
Figure \ref{fig.febiasbig}. At these large radii the 1T, 2T, and PLDEM
models fit quite similarly: $\chi^2=891.9$ for 762 dof (1T),
$\chi^2=840.0$ for 760 dof (2T), and $\chi^2=842.3$ for 760 dof
(PLDEM). We do not see the residuals near 1~keV characteristic of the
Fe Bias. This is expected since the effective range of temperatures
within this aperture is only from $T\approx 1.15-1.25$~keV according
to the spatially resolved analysis (see in particular Figure 3 of
Paper 1). This small temperature range does not fully span the Fe L
region and will not suffer a substantial Fe Bias.

In Table \ref{tab.abun3} we present the results for the iron abundance
and abundance ratios and the estimated error budget analogously to
that for the central region in the previous section.  Although the
spatially resolved analysis implies a small range of temperatures, we
do consider the 1T results in our error budget for the ``Model''
simply to be conservative since the $\chi^2$ fits do not obviously
distinguish the multitemperature models from the 1T models.  (We
mention that the 2T model gives $\thot=1.33\pm 0.02\solar$ and
$\tcool=0.78\pm 0.08\solar$ with a best-fitting $norm_{\rm
h}/norm_{\rm c} = 5.8$. The PLDEM model gives $\alpha= 4.5\pm 0.4$,
$\tmin = 0.29\pm 0.13$~keV, $\tmax-\tmin= 1.12\pm 0.11$~keV. The DEMs
of these models are very peaked around a single temperature $\approx
1.35$~keV; e.g., the multiphase strength \citep{buot99b} of the
best-fitting PLDEM model is $\sigma_{\xi}\approx 0.10$ which indicates
a nearly single-temperature medium.)

The value of \fe\ is tightly constrained to $\approx 40\%$ of the
average value of \fe\ within $R=5\arcmin$. Notice also that the
\si/\fe\ and \su/\fe\ ratios are consistent with their corresponding
values for $R=5\arcmin$ within their $1-1.5\sigma$ statistical
errors. The Mg and Ne abundances have rather large systematic errors
in this outer region, and thus must be considered to be consistent
with their values for $R=5\arcmin$. However, \ox/\fe\ is significantly
larger and has a value near solar. As shown in \S \ref{other}, the
large value of \ox\ at large radius is due entirely to bin \#7 which
suggests it may not be that large. Values of \ox/\fe\ at adjacent
radii are consistent with those at smaller radii within the $\approx
1\sigma$ errors. If the discrepancy is real, it is difficult to
understand why \ox/\fe\ increases with radius while the other ratios
are consistent with a constant or decreasing profile. Alternatively,
there could be a calibration issue peculiar to the measurement of the
oxygen abundance.

\section{Conclusions}
\label{conc}

\subsection{Iron Abundance and Bias}

One robust conclusion to be drawn from our spatially resolved spectral
analysis of the \xmm\ data of NGC 5044 is that the iron abundance
drops from a value near solar at $r\approx 30$~kpc to a value between
0.3-0.4 solar at the largest radius probed ($\approx 100$~kpc). This
radial decrease in \fe\ is highly significant considering both the
statistical and systematic errors as illustrated by the
emission-weighted average abundances obtained from two large apertures
(\S \ref{bias}): $\fe=1.09\pm 0.04 \solar\, (\rm statistical)\, \pm
0.05\solar\, + 0.18\solar\, (\rm systematic)$ for $R=5\arcmin$
compared to $\fe=0.44\pm 0.02 \solar\, (\rm statistical)\, \pm
0.10\solar\, + 0.13\solar\, (\rm systematic)$ for
$R=5\arcmin-10\arcmin$. (Note that for ease of presentation we have
merely added the systematic errors listed in Table
\ref{tab.abun2}. The reader should refer to that table for the explicit
breakdown of the errors.)  There is no evidence that the radial
decline of \fe\ flattens at large radius.  The gradient in \fe\ at
large radius measured by \xmm\ is more precise than obtained from
previous \rosat\ and \asca\ studies of NGC 5044
\citep[e.g.,][]{davi94,fino99,buot00c}.  We quote values of
\fe\ with respect to the new solar photospheric value
\citep[e.g.,][]{mcwi97,grsa} which also agrees with the meteoritic
value.

A second robust conclusion is that the values of \fe\ within the
central regions ($r\la 30$~kpc) are not highly sub-solar; i.e., not
less than $0.5\solar$. In contrast, we find that simultaneous spectral
fits to the \xmm\ and \chandra\ data within the central regions give
values of $\fe\approx 1\solar$ for all deprojected temperature models
assuming only foreground Galactic absorption. For the preferred
multitemperature models (2T and PLDEM) we obtain $\fe\approx
1.2\solar$ over $5-30$~kpc which dips to $\fe\approx 0.8\solar$ for
$r<5$~kpc. (These values of \fe\ obtained from multitemperature models
assume that the abundances are the same for each temperature
component.) The emission-weighted average value of \fe\ quoted above
for $R=5\arcmin$ (48~kpc) rules out with high significance a very
sub-solar average value of \fe.  We conclude that the very sub-solar
Fe values obtained from previous analyses of \asca\ data in the
central region of NGC 5044 and related systems were affected by the Fe
Bias as argued previously by us
\citep[][]{buot98c,buot99a,buot00a,buot00c} and others
\citep{alle00,mole01a}. Recently, we have obtained a similar result
for the galaxy group NGC 1399 with \xmm\ data \citep{buot02a}.

We emphasize that for moderate resolution spectra, such as provided by
the CCDs of the \xmm\ and \chandra\ satellites, the Fe abundance of
$\sim 1$~keV systems like NGC 5044 is not determined from a single
isolated emission line, but rather from the unresolved Fe L shell
emission lines that dominate the spectrum between $\approx
0.7-1.2$~keV. Consequently, no single Fe emission line is responsible
for the Fe abundance inferred from current CCD studies.  Hence, while
20\%-30\% errors in some Fe L lines still exist in the
\apec\ and \mekal\ plasma codes \citep{beha01}, errors of this
magnitude cannot be responsible for large systematic errors in the
inferred Fe abundances from the moderate-resolution CCD data. The
small (5\%-10\%) systematic offset in the Fe abundances deduced from
the \apec\ and \mekal\ codes for NGC 5044 (\S \ref{plasma}), each of
which has different line libraries and approximation methods, indicates
the magnitude of the systematic error in the Fe abundance expected
from remaining errors in the modeling of the Fe L lines.

Of more serious concern is that Fe abundance measurements can be
systematically biased by improper definition of the continuum,
inclusion of an intrinsic absorber component when none exists, and
forcing a single-temperature component to fit a multitemperature
spectrum possessing temperature components near 1~keV. So long as the
lower energy range used for spectral fitting is well below the Fe L
region (i.e., at least as low as 0.5~keV), the bias arising from a
poor continuum definition is mitigated (\S \ref{emin}). Since 1T
models with intrinsic absorption do not fit as well as 2T and PLDEM
models with Galactic \nh, and there is no evidence outside the X-ray
band for large quantities of cold (or warm) absorbing gas in NGC 5044
or other X-ray luminous groups and clusters, it is reasonable to
include only foreground absorption from the Milky Way (\S
\ref{nh}). Finally, the Fe bias associated with forcing a
single-temperature component to fit the multitemperature spectra in
the central regions of NGC 5044 results in underestimates of \fe\
often by as much as 50\% or more.

\subsection{Implications for Supernova Enrichment}

\subsubsection{Iron}

The approximately solar values for \fe\ within the central $\sim
30$~kpc determined from our X-ray observations of the hot gas exceed
the typical values of $\approx 0.5\solar$ obtained for the stars
averaged over a region of half an effective radius \citep{trag00a}.
This result implies that the hot gas in NGC 5044 could have been
enriched by a substantial amount of Type Ia supernovae (\snia) which
alleviates the most serious conflict of previous X-ray observations of
elliptical galaxies and centrally E-dominated groups with the models
of chemical enrichment that assume a Milky Way IMF and \snia\ rate as
observed in elliptical galaxies \citep{capp02}.  

This conflict, called the ``Iron Discrepancy'' in elliptical galaxies
by \citet{arim97} and \citet{renz97}, is partially resolved for NGC
5044 by our \xmm\ and \chandra\ observations which demonstrate that
previous underestimates of the iron abundances with \asca\ arose
partially from the Fe Bias and partially from the use of the incorrect
solar abundance standard. Our recent \xmm\ observation of the
centrally E-dominated group NGC 1399 \citep{buot02a} provides similar
results for the iron abundance. Other \xmm\ observations (EPIC and
RGS) of dominant ellipticals in groups (e.g., NGC 533,
\citealt{pete03}; NGC 4636, \citealt{xu02a}) and clusters (e.g., M87,
\citealt{mole01a,gast02a}; A2029, \citealt{lewi02a}) find near solar
central Fe abundances consistent with significant \snia\ enrichment.

\subsubsection{$\alpha$/Fe Ratios}

The ratios of the abundances of the elements O, Mg, Si, and S to Fe
provide strong constraints on the fraction of iron in the hot gas
produced by \snia. As we discussed in \S \ref{si} and \S \ref{bias},
next to Fe the best constrained abundance is that of Si. The
emission-weighted average \si/\fe\ ratio within $R=5\arcmin$ (48~kpc)
is, $\si/\fe = 0.832 \pm 0.017\, (\rm statistical)\, \pm 0.018\,
+0.071\, (\rm systematic)$ in solar units. This ratio translates to a
\snia\ fraction of 67\%-79\% where we have used (1) the
$1\sigma$ statistical error on $\si/\fe$ , (2) the theoretical
\snia\ yield of the ``convective deflagration'' W7 model of
\citet{nomo97b}, and (3) the range of theoretical \snii\ yields
reported in \citet{gibs97}. Note that for this calculation we have
used the abundance ratios for the theoretical supernova models
expressed in terms of the \citet{grsa} solar abundance standard as
reported in Table 2 of \citet{gast02a}. Using the $1\sigma$ range for
\si/\fe\ at large radius (i.e., for $R=46-96$~kpc) we obtain a \snia\
fraction of 70\%-86\%, consistent with the result within $R=48$~kpc.

The emission-weighted average sulfur-to-iron abundance ratio within
$R=5\arcmin$ (48~kpc) is, $\su/\fe = 0.542 \pm 0.020\, (\rm
statistical)\, \pm 0.010\, +0.007\, (\rm systematic)$ in solar units,
which is consistent with the average value obtained at larger radii
out to 100~kpc. Using the $1\sigma$ statistical error on the X-ray
measurement, along with the theoretical SNe models mentioned above, we
obtain a \snia\ fraction of 52\%-82\%, in good agreement with that
inferred from the \si/\fe\ ratio.

The largest contributor to the uncertainty in the \snia\ fraction
deduced from these abundance ratios is the theoretical SNe yields. For
example, if we assume the \snii\ yields of \citet{nomo97a} we obtain
\snia\ fractions of 76\%-77\% from \si/\fe\ and 70\%-75\% from
\su/\fe\ using the $1\sigma$ statistical errors on the X-ray
measurements within $R=48$~kpc.  The so-called ``delayed-detonation''
\snia\ models, WDD1 and WDD2, of \citet{nomo97b} appear to be
incompatible with the measured \si/\fe\ and \su/\fe\ ratios. The WDD1
model predicts $\si/\fe=1.15$ and $\su/\fe=1.01$ in solar units which
are incompatible with our measurements. The WDD2 model predicts
$\su/\fe=0.62$ which is $4\sigma$ above the measured value within
$R=48$~kpc. However, if the various systematic errors shown in Table
\ref{tab.abun2} are considered then the discrepancy is not highly
significant. Nevertheless, the \si/\fe, and especially \su/\fe, ratios
appear to prefer the convective deflagration W7 \snia\ model over the
delayed-detonation models.

Hence, we infer that \snia\ have contributed $70\%-80\%$ of the iron
mass within a 100~kpc radius of NGC 5044 -- a result that is fully
consistent with that obtained from our recent analysis of an
\xmm\ observation of the bright nearby galaxy group, NGC 1399
\citep{buot02a}. This \snia\ fraction is also similar to that inferred
for the Sun and therefore suggests a stellar initial mass function
similar to the Milky Way \citep[e.g.,][]{renz93,renz97,wyse97}.

We mention that the uncertain Ne abundance is also consistent with
this result -- see \S \ref{bias}. The average \ox/\fe\ ratio (except
bin \#7) is also consistent with this level of \snia\ enrichment:
75\%-89\% using the $1\sigma$ statistical error on the X-ray
measurement within $R=48$~kpc, the W7 \snia\ model, and the range in
\snii\ yields from \citet{gibs97}. Like \ox/\fe\ in bin \#7, the
near-solar Mg/Fe ratio in the central regions implies a much smaller
\snia\ fraction ($\approx 50\%$).  We cannot explain the apparent
anomalous behavior of these \ox/\fe\ and \mg/\fe\ ratios, but since
their relevant lines are blended with strong Fe L lines and (in the
case of Mg) calibration lines, we consider these discrepancies to be
tentative.

\subsection{Central Iron Deficit and Problems with Enrichment Models}

\begin{figure*}[t]
\centerline{\psfig{figure=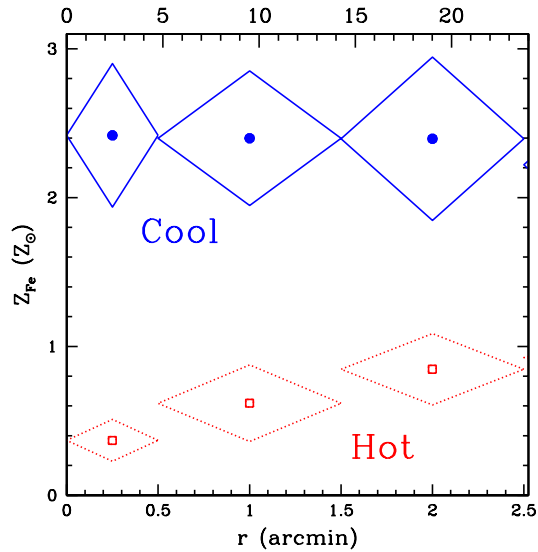,height=0.3\textheight}}
\caption{\label{fig.diff} Radial profiles
(units -- bottom: arcminutes, top: kpc) of the Fe abundance and
associated $1\sigma$ errors for the 2T (2D) model where the abundances
have been allowed to vary separately for each component. Only shown
are the the results for the inner shells.}
\end{figure*} 

Although the near-solar central value of the iron abundance and the
values of the $\alpha$/Fe ratios provide strong evidence for \snia\
enrichment of the hot gas from the central galaxy, there remain some
problems with this scenario. First, the abundances in the hot gas
generally rise toward the center and then dip in the innermost bin;
this dip is more pronounced for the deprojected abundance
profiles. Such a dip is difficult to understand in the context of
enrichment from the central galaxy which should be most pronounced at
the very center. Some cD clusters also show such a dip \citep[e.g.,
Centaurus,][]{sand02a} which \citet{morr02a} have suggested is an
artifact of attempting to model a highly inhomogeneous metal
distribution with a homogeneous spectral model. This model is
attractive since it could explain why sub-keV cooling gas has not been
found by \chandra\ or \xmm\ in cooling flows.

In NGC 5044 we find that the central abundance dip is indeed sensitive
to assumptions in the spectral model. (Central dips are also sensitive
to the lower energy limit of the bandpass as discussed in \S
\ref{emin}.)  If we tie oxygen to iron in their solar ratio the dip in
the abundances is less pronounced. In the spirit of \citet{morr02a},
if instead we allow the abundances on each temperature component of
the 2T model to vary separately then the abundances on the cool
component remain large all the way to the center. We show the central
iron abundance profile of this model in Figure \ref{fig.diff}. Because
of the large number of free parameters in this model the precise
radial variation of \fe\ in Figure \ref{fig.diff} should be treated
with caution. But the basic idea that the property of a central dip is
sensitive to further assumptions about the spectral model is clear.

Second, we note that although correcting for the Fe Bias partially
removes the ``Iron Discrepancy'' noted by \citet{arim97}, chemical
models of elliptical galaxies without cooling flows predict central
iron abundances even larger than we have measured for N5044 and NGC
1399 \citep{brig03a}.

It should also be emphasized that the properties of the central $\sim
5$ kpc (i.e., central $R\approx 30\arcsec$) are distinctly different
from those at immediately larger radii. Hence, studies of NGC 5044
with the high-resolution gratings of \chandra\ and \xmm\ can only
probe this special region and cannot tell us about the properties at
larger radii. We note, however, that the RGS, EPIC, and ACIS-S3 data
give consistent results in their overlap regions (\S
\ref{calib}).

\acknowledgements It is a pleasure to thank T. Fang for communicating
results of his unpublished analysis of the RGS data of NGC 5044.  We
gratefully acknowledge partial support from NASA grants NAG5-9956,
NAG5-10758, and NAG5-10748.

\bibliographystyle{apj}

\begin{thebibliography}{44}
\expandafter\ifx\csname natexlab\endcsname\relax\def\natexlab#1{#1}\fi

\bibitem[{{Allen} {et~al.}(2000){Allen}, {Di Matteo}, \& {Fabian}}]{alle00}
{Allen}, S.~W., {Di Matteo}, T., \& {Fabian}, A.~C. 2000, \mnras, 311, 493

\bibitem[{{Anders} \& {Grevesse}(1989)}]{angr}
{Anders}, E. \& {Grevesse}, N. 1989, \gca, 53, 197

\bibitem[{{Arimoto} {et~al.}(1997){Arimoto}, {Matsushita}, {Ishimaru},
  {Ohashi}, \& {Renzini}}]{arim97}
{Arimoto}, N., {Matsushita}, K., {Ishimaru}, Y., {Ohashi}, T., \& {Renzini}, A.
  1997, \apj, 477, 128

\bibitem[{{Awaki} {et~al.}(1994){Awaki}, {Mushotzky}, {Tsuru}, {Fabian},
  {Fukazawa}, {Loewenstein}, {Makishima}, {Matsumoto}, {Matsushita}, {Mihara},
  {Ohashi}, {Ricker}, {Serlemitsos}, {Tsusaka}, \& {Yamazaki}}]{awak94}
{Awaki}, H., {Mushotzky}, R., {Tsuru}, T., {Fabian}, A.~C., {Fukazawa}, Y.,
  {Loewenstein}, M., {Makishima}, K., {Matsumoto}, H., {Matsushita}, K.,
  {Mihara}, T., {Ohashi}, T., {Ricker}, G.~R., {Serlemitsos}, P.~J., {Tsusaka},
  Y., \& {Yamazaki}, T. 1994, \pasj, 46, L65

\bibitem[{{Behar} {et~al.}(2001){Behar}, {Cottam}, \& {Kahn}}]{beha01}
{Behar}, E., {Cottam}, J., \& {Kahn}, S.~M. 2001, \apj, 548, 966

\bibitem[{{Brighenti} \& {Mathews}(1999)}]{brig99a}
{Brighenti}, F. \& {Mathews}, W.~G. 1999, \apj, 515, 542

\bibitem[{{Brighenti} \& {Mathews}(2003)}]{brig03a}
---. 2003, \apj, 587, 580

\bibitem[{{Buote}(1999)}]{buot99a}
{Buote}, D.~A. 1999, \mnras, 309, 685

\bibitem[{{Buote}(2000{\natexlab{a}})}]{buot00c}
---. 2000{\natexlab{a}}, \apj, 539, 172

\bibitem[{{Buote}(2000{\natexlab{b}})}]{buot00a}
---. 2000{\natexlab{b}}, \mnras, 311, 176

\bibitem[{{Buote}(2002)}]{buot02a}
---. 2002, \apjl, 574, L135

\bibitem[{{Buote} {et~al.}(1999){Buote}, {Canizares}, \& {Fabian}}]{buot99b}
{Buote}, D.~A., {Canizares}, C.~R., \& {Fabian}, A.~C. 1999, \mnras, 310, 483

\bibitem[{{Buote} \& {Fabian}(1998)}]{buot98c}
{Buote}, D.~A. \& {Fabian}, A.~C. 1998, \mnras, 296, 977

\bibitem[{{Buote} {et~al.}(2003){Buote}, {Lewis}, {Brighenti}, \&
  {Mathews}}]{buot03a}
{Buote}, D.~A., {Lewis}, A.~D., {Brighenti}, F., \& {Mathews}, W.~G. 2003,
  \apj, in press (astro-ph/0205362)

\bibitem[{{Cappellaro} \& {Turatto}(2002)}]{capp02}
{Cappellaro}, E. \& {Turatto}, M. 2002, in The influence of binaries on stellar
  population studies ed. D. Vanbeveren (Brussels 21-25 Aug. 2000), in press
  (astro-ph/0012455)

\bibitem[{{David} {et~al.}(1994){David}, {Jones}, {Forman}, \&
  {Daines}}]{davi94}
{David}, L.~P., {Jones}, C., {Forman}, W., \& {Daines}, S. 1994, \apj, 428, 544

\bibitem[{{Davis} {et~al.}(1999){Davis}, {Mulchaey}, \& {Mushotzky}}]{davi99}
{Davis}, D.~S., {Mulchaey}, J.~S., \& {Mushotzky}, R.~F. 1999, \apj, 511, 34

\bibitem[{{Finoguenov} \& {Ponman}(1999)}]{fino99}
{Finoguenov}, A. \& {Ponman}, T.~J. 1999, \mnras, 305, 325

\bibitem[{{Fukazawa} {et~al.}(1996){Fukazawa}, {Makishima}, {Matsushita},
  {Yamasaki}, {Ohashi}, {Mushotzky}, {Sakima}, {Tsusaka}, \&
  {Yamashita}}]{fuka96}
{Fukazawa}, Y., {Makishima}, K., {Matsushita}, K., {Yamasaki}, N., {Ohashi},
  T., {Mushotzky}, R.~F., {Sakima}, Y., {Tsusaka}, Y., \& {Yamashita}, K. 1996,
  \pasj, 48, 395

\bibitem[{{Gastaldello} \& {Molendi}(2002)}]{gast02a}
{Gastaldello}, F. \& {Molendi}, S. 2002, \apj, 572, 160

\bibitem[{{Gibson} {et~al.}(1997){Gibson}, {Loewenstein}, \&
  {Mushotzky}}]{gibs97}
{Gibson}, B.~K., {Loewenstein}, M., \& {Mushotzky}, R.~F. 1997, \mnras, 290,
  623

\bibitem[{{Grevesse} \& {Sauval}(1998)}]{grsa}
{Grevesse}, N. \& {Sauval}, A.~J. 1998, Space Science Reviews, 85, 161

\bibitem[{{Lewis} {et~al.}(2002){Lewis}, {Stocke}, \& {Buote}}]{lewi02a}
{Lewis}, A.~D., {Stocke}, J.~T., \& {Buote}, D.~A. 2002, \apjl, 573, L13

\bibitem[{{Loewenstein} \& {Mushotzky}(2002)}]{loew02}
{Loewenstein}, M. \& {Mushotzky}, R.~F. 2002, in IAU Symposium, 139

\bibitem[{{Matsumoto} {et~al.}(1997){Matsumoto}, {Koyama}, {Awaki}, {Tsuru},
  {Loewenstein}, \& {Matsushita}}]{mats97}
{Matsumoto}, H., {Koyama}, K., {Awaki}, H., {Tsuru}, T., {Loewenstein}, M., \&
  {Matsushita}, K. 1997, \apj, 482, 133

\bibitem[{{Matsushita} {et~al.}(1994){Matsushita}, {Makishima}, {Awaki},
  {Canizares}, {Fabian}, {Fukazawa}, {Loewenstein}, {Matsumoto}, {Mihara},
  {Mushotzky}, {Ohashi}, {Ricker}, {Serlemitsos}, {Tsuru}, {Tsusaka}, \&
  {Yamazaki}}]{mats94}
{Matsushita}, K., {Makishima}, K., {Awaki}, H., {Canizares}, C.~R., {Fabian},
  A.~C., {Fukazawa}, Y., {Loewenstein}, M., {Matsumoto}, H., {Mihara}, T.,
  {Mushotzky}, R.~F., {Ohashi}, T., {Ricker}, G.~R., {Serlemitsos}, P.~J.,
  {Tsuru}, T., {Tsusaka}, Y., \& {Yamazaki}, T. 1994, \apjl, 436, L41

\bibitem[{{Matsushita} {et~al.}(2000){Matsushita}, {Ohashi}, \&
  {Makishima}}]{mats00}
{Matsushita}, K., {Ohashi}, T., \& {Makishima}, K. 2000, \pasj, 52, 685

\bibitem[{{McWilliam}(1997)}]{mcwi97}
{McWilliam}, A. 1997, \araa, 35, 503

\bibitem[{{Molendi} \& {Gastaldello}(2001)}]{mole01a}
{Molendi}, S. \& {Gastaldello}, F. 2001, \aap, 375, L14

\bibitem[{{Morris} \& {Fabian}(2002)}]{morr02a}
{Morris}, R.~G. \& {Fabian}, A.~C. 2002, in ASP Conf. Ser. 253: Chemical
  Enrichment of Intracluster and Intergalactic Medium, 85--+

\bibitem[{{Mulchaey}(2000)}]{mulc00}
{Mulchaey}, J.~S. 2000, \araa, 38, 289

\bibitem[{{Mushotzky} {et~al.}(2003){Mushotzky}, {Figueroa-Feliciano},
  {Loewenstein}, \& {Snowden}}]{mush03a}
{Mushotzky}, R., {Figueroa-Feliciano}, E., {Loewenstein}, M., \& {Snowden},
  S.~L. 2003, (astro-ph/0302267)

\bibitem[{{Nomoto} {et~al.}(1997{\natexlab{a}}){Nomoto}, {Hashimoto}, \&
  {Tsujimoto}}]{nomo97a}
{Nomoto}, K., {Hashimoto}, M., \& {Tsujimoto}, T. 1997{\natexlab{a}}, \nphysa,
  616, 79

\bibitem[{{Nomoto} {et~al.}(1997{\natexlab{b}}){Nomoto}, {Iwamoto}, {Nakasoto},
  {Thielemann}, {Brachwitz}, {Tsujimoto}, {Kubo}, \& {Kishimoto}}]{nomo97b}
{Nomoto}, K., {Iwamoto}, K., {Nakasoto}, N., {Thielemann}, F.~K., {Brachwitz},
  F., {Tsujimoto}, T., {Kubo}, Y., \& {Kishimoto}, N. 1997{\natexlab{b}},
  \nphysa, 621, 467

\bibitem[{{Peterson} \& {et.~al.}(2003)}]{pete03}
{Peterson}, J.~R. \& {et.~al.} 2003, \apj, submitted (astro-ph/0210662)

\bibitem[{{Renzini}(1997)}]{renz97}
{Renzini}, A. 1997, \apj, 488, 35

\bibitem[{{Renzini}(2000)}]{renz00}
{Renzini}, A. 2000, in Large Scale Structure in the X-ray Universe, Proceedings
  of the 20-22 September 1999 Workshop, Santorini, Greece, eds. Plionis, M. \&
  Georgantopoulos, I., Atlantisciences, Paris, France, p.103, 103

\bibitem[{{Renzini} {et~al.}(1993){Renzini}, {Ciotti}, {D'Ercole}, \&
  {Pellegrini}}]{renz93}
{Renzini}, A., {Ciotti}, L., {D'Ercole}, A., \& {Pellegrini}, S. 1993, \apj,
  419, 52

\bibitem[{{Sanders} \& {Fabian}(2002)}]{sand02a}
{Sanders}, J.~S. \& {Fabian}, A.~C. 2002, \mnras, 331, 273

\bibitem[{{Tamura} {et~al.}(2003){Tamura}, {Kaastra}, {Makishima}, \&
  {Takahashi}}]{tamu03a}
{Tamura}, T., {Kaastra}, J.~S., {Makishima}, K., \& {Takahashi}, I. 2003, \aap,
  399, 497

\bibitem[{{Tonry} {et~al.}(2001){Tonry}, {Dressler}, {Blakeslee}, {Ajhar},
  {Fletcher}, {Luppino}, {Metzger}, \& {Moore}}]{tonr01}
{Tonry}, J.~L., {Dressler}, A., {Blakeslee}, J.~P., {Ajhar}, E.~A., {Fletcher},
  A.~., {Luppino}, G.~A., {Metzger}, M.~R., \& {Moore}, C.~B. 2001, \apj, 546,
  681

\bibitem[{{Trager} {et~al.}(2000){Trager}, {Faber}, {Worthey}, \& {Gonz{\'
  a}lez}}]{trag00a}
{Trager}, S.~C., {Faber}, S.~M., {Worthey}, G., \& {Gonz{\' a}lez}, J.~J.~.
  2000, \aj, 119, 1645

\bibitem[{{Wyse}(1997)}]{wyse97}
{Wyse}, R.~F.~G. 1997, \apjl, 490, L69

\bibitem[{{Xu} {et~al.}(2002){Xu}, {Kahn}, {Peterson}, {Behar}, {Paerels},
  {Mushotzky}, {Jernigan}, {Brinkman}, \& {Makishima}}]{xu02a}
{Xu}, H., {Kahn}, S.~M., {Peterson}, J.~R., {Behar}, E., {Paerels}, F.~B.~S.,
  {Mushotzky}, R.~F., {Jernigan}, J.~G., {Brinkman}, A.~C., \& {Makishima}, K.
  2002, \apj, 579, 600

\end{thebibliography}

\end{document}